\documentclass[a4paper]{article}

\usepackage[utf8]{inputenc}
\usepackage{amsmath}
\usepackage{amsfonts}
\usepackage{amssymb}
\usepackage{longtable}
\usepackage{floatpag}
\usepackage{bbold}

\usepackage[makeroom]{cancel}

\usepackage{appendix}
\usepackage{todonotes}
\title{Calculation of Epidemic Arrival Time Distributions using Branching processes}
\author{Alastair Jamieson-Lane, Bernd Blasius }
\date{March 2020}

\usepackage{graphicx}

\begin{document}

\maketitle
\begin{abstract}
The rise of the World Airline Network over the past century has lead to
sharp changes in our notions of ‘distance’ and ‘closeness’ - both in terms
of trade and travel, but also (less desirably) with respect to the spread of
disease. When novel pathogens are discovered, countries, cities and hospitals are caught trying to predict how much time they have to prepare.
In this paper, by considering the early stages of epidemic spread as a simple branching process, we derive the full probability distribution of arrival times. We are able to re-derive a number of past arrival time results (in suitable limits), and demonstrate the robustness of our approach, both to parameter values far outside the traditionally considered regime, and to errors in the parameter values used. The branching process approach provides some theoretical justification to the ‘effective distance’ introduced by Brockmann \& Helbing (2013), however we also observe that when compared to real world data, the predictive power of all methods in this class is significantly lower than has been previously reported.
\end{abstract}

\section{Introduction}

It was said by Jules Verne in 1873 that ``The world has grown smaller, since a man can now go round it ten times more quickly than a hundred years ago,'' that one could travel around the world in eighty days \cite{verne_around_1873}.
One can't help but wonder what Verne would think of the advances in technology that have occurred in the century and a half since, which now allow the circumnavigation of the globe in less than eighty hours.
The World Aviation Network (WAN) has, over the past century, shrunk the globe to a fraction of its former size, allowing for vast increases in tourism, immigration, and trade.
At the same time, diseases that might once have travelled at the speed of cart, ship or train now cross the world in a number of days or hours. In 2003 Severe Acute Respiratory Syndrome (SARS) was able to spread from China to Vietnam, Hong Kong and Canada within two weeks of being reported to the World Health Organization \cite{who_who_2003}. Similarly, the 2009 H1N1 flu pandemic first reported in Mexico was able to reach both Europe and Asia within a fortnight \cite{who_who_2009}. Now, in 2020 we see that  coronavirus SARS-CoV-2 has spread quickly across the world \cite{who_novel_2020}, reaching all but a handful of countries withing a few short months.

Modelling of disease spread and disease arrival time is critical, both to mitigate the effects of a disease and to determine where best to apply quarantine and screening protocols so as to minimize the spread of infection.
In order to model the spread of disease, given the current WAN, a number of highly detailed epidemic simulation models have been created, such as GLEAMviz\cite{broeck_gleamviz_2011}. Depending on how they are set up, these models can include such factors as vaccination, incubation times, multiple susceptibility classes, non-linear responses, seasonal forcing, quarantine, and the stochastic movement of individual agents, providing a detailed and comprehensive modelling framework. Unfortunately, in many cases, such models quickly become black boxes in and of themselves - objects to be studied and analyzed only via costly simulation - allowing predictions of the future that can be observed, but seldom understood. 
This difficulty is further exacerbated by the difficulties of determining parameter values, often a delicate task during the initial stages of a disease outbreak when information is scarce.

Depending on the questions being asked, explicit modelling of such intricate details is necessary, however, for other questions, such as the determination of epidemic arrival time (`AT'), it seems that simpler methods may suffice.
For this reason, a number of authors \cite{brockmann_hidden_2013,iannelli_effective_2017, gautreau_arrival_2007, chen_estimating_2018} have proposed a variety of heuristics and metrics -- artificial measures of distance based on flight data from the WAN. The goal of such metrics is to predict relative arrival times for an epidemic starting in a specified location, predicting for example that an epidemic starting in Vancouver will take twice as long to reach Istanbul as it will to reach Rio de Janeiro. Absolute AT will also depend on the parameters of the particular disease in question.

A particularly elegant example of such a metric is Brockmann \& Helbing's \cite{brockmann_hidden_2013} `effective distance', in which they define the direct distance between a pair of locations as 
\begin{equation}
    d_{ij} = 1 - \log(P_{ij}),
    \label{Eq:ED}
\end{equation}
where $P_{ij}$ is the probability that a particular individual who is leaving location $i$ is travelling to location $j$. 
The total effective distance between nodes $a$ and $b$ is then given as the shortest path between a given pair of nodes:
\begin{equation}
    D_{ab} = \min \sum_{ij} d_{ij} = \min \sum_{ij} \left[ 1 - \log(P_{ij}) \right],
    \label{eq:OriBrock}
\end{equation}
where here we minimize over all possible paths from $a$ to $b$, and $d_{ij}$ denotes the effective distance along each step of this shortest path (and $P_{ij}$ the corresponding probability). 

Papers by Gautreau, Ianelli, et al. provide more detailed and mathematically justified metrics, by first calculating the expected time for a growing epidemic to make its first `jump' \cite{gautreau_arrival_2007}, and then summing over all possible paths joining a pair of airports (in an appropriate manner) \cite{iannelli_effective_2017}.
More recent work by Chen et al.
\cite{chen_estimating_2018} made use of linear spreading theory and the matrix exponential, in order to attain similar accuracy at reduced computational cost.% to calculate the time when the expected number of infections is equal to one, a quantity that is highly correlated with, but distinct from, the expected time when the actual number of infections is one. They achieve the same high accuracy as Ianelli et al. when compared to simulations, but at significantly lower computational cost.

Underlying each of the above efforts, either implicitly or explicitly, is an assumption of unbound growth; the idea that in the early stages of an epidemic, disease spread is not limited by population size. Such unbound growth can, and often has been, modeled as smooth exponential growth, a continuous approximation of discrete real world populations. An alternative model for unbound growth, one often used when small populations, rare events, and probabilities are of interest, is the branching process \cite{kimmel_branching_2015}. The branching process is a classical tool in the study of extinction and evolution\cite{schaffer_survival_1970}, and has been used to study epidemics\cite{gonzalez_workshop_2010}, surnames and genealogies \cite{watson_probability_1875}, and cancers \cite{goldie_mathematic_1979,coldman_stochastic_1986}. %A rather nice review of Branching Processes and their historical contexts is provided by Gonz\'alez and Puerto \cite{gonzalez_branching_2010}. 
In this paper we apply the branching process framework in order to calculate not only the mean, but also the full distribution of ATs for arbitrary networks.

In section \ref{sect:model} we introduce a basic `multi-compartment branching process' model, and from this model derive a system of ODE's that precisely determine the probability of epidemic arrival by a given time. In section
\ref{sect:meanVar} we determine the mean and variance in the AT, and explore how these results both support and contrast with the results of past papers.
Section \ref{sect:limits} explores the sensitivity of our predictions to perturbations to system parameters and network structure.
In section \ref{sect:world} we use data from real world flight networks, and compare our predictions, along with the predictions of past authors to epidemic arrival times as observed for in the 2003 SARS epidemic, and 2009 H1N1 influenza pandemic. In both cases we observe correlation between predicted and observed results, but also significant noise. We are, unfortunately, unable to reproduce previously published results, and instead observe the observed stochasticity is both larger than previously reported, and larger than might be expected given the underlying model.

Our goal in this paper is to provide a `canonical' approach to calculating arrival times; that is to say, an approach which relies on minimal assumptions and approximations, and can be trusted to provide accurate results across all of parameter space. To the extent that the method works, it can be seen as providing a foundation to past methods and heuristics. To the extent that our predictions disagree with real world data this is suggestive of either flaws in the data, or gaps in the underlying model, gaps that will require not simply better mathematics, but instead better understanding of the system under study.

\section{Diffusion on a Network and the Branching Process}
\label{sect:model}
Let us begin by concretely defining the model; we consider a network of $N$ connected nodes, each node representing a location. In our case these nodes refer to particular airports in the WAN. Individuals travel from node $i$ to node $j$ at transport rate $\gamma T_{ij}$. Here $T_{ij}$ gives the relative transport rate along a given route, while $\gamma$ is the global flight rate. While $\gamma$ and $T$ can be folded together as one parameter, maintaining this distinction will prove convenient later. We select $T_{ii}$ to represent the rate at which individuals \emph{leave} location $i$; $T_{ii} = -\sum_{j\neq i} T_{ij}$.  Each node contains a population of $I_k(t)$ infected individuals, initially set to zero. This population evolves according to a continuous time branching process: in any given small time interval $dt$ $I_k(t)$ will increase by one with probability $\alpha_k I_k(t) dt$ and decrease by one with probability $\beta_k I_k(t) dt$. This represents infection and recovery ($\alpha$ and $\beta$ respectively). Parallel with this branching process, infected individuals may travel from one location, $i$, to a neighbouring location, $j$, with probability $\gamma T_{ij} I_i(t) dt$, resulting in an increment at location $j$ and a decrement at location $i$. We thus have what might be described as a ``continuous time multitype branching process'' \cite{kimmel_branching_2015}.

\begin{figure}[h]
    \centering
\centerline{\includegraphics[width=1.05\textwidth]{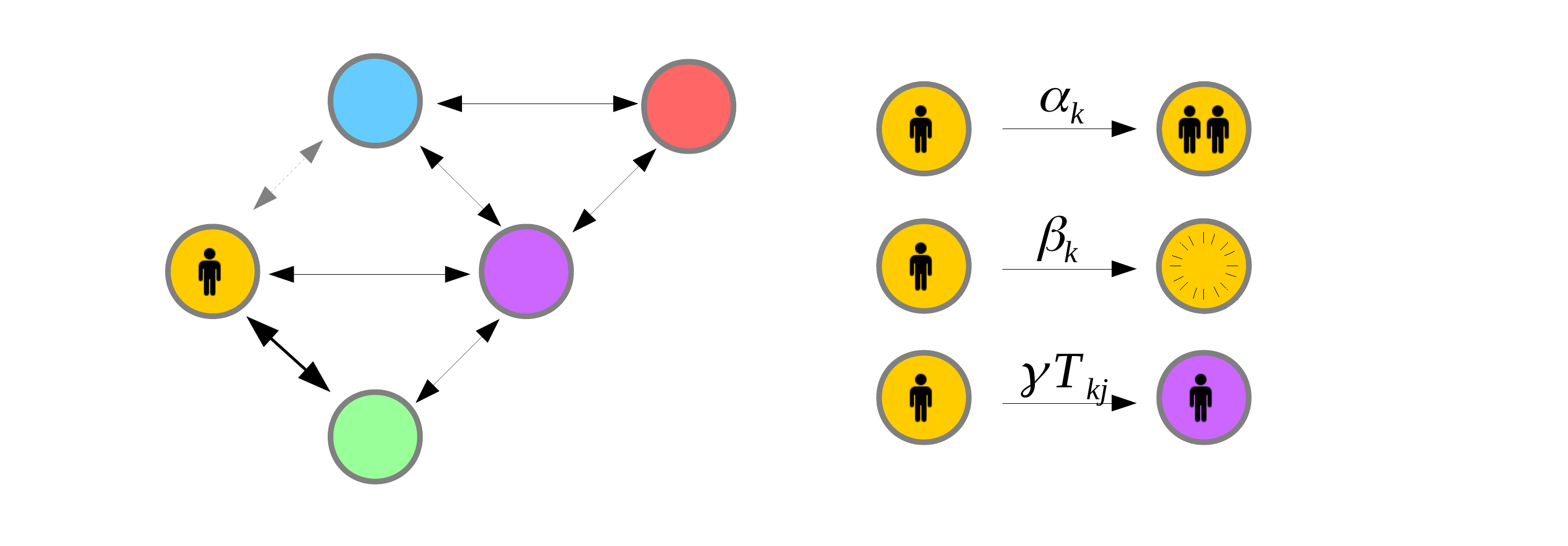}}
    \caption{(Left) Schematic diagram of a traveller in a network; each node representing a city/airport, and edges representing a flight link in the WAN. Edges will have different weights depending on the number of passengers flying. (Right) Each infected individual in location $k$ can take one of three actions at any given moment of time: they may infected another individual (at infection rate $\alpha_k$), they may recover, removing themselves from the infectious population (at recovery rate $\beta_k$), or they may travel to some other location $j$ at transport rate $\gamma T_{kj}$.
    \label{fig:NetworkReactCartoon}}
\end{figure}
We would like to determine the travel time from $a$ to $b$. Stated formally, suppose a novel diseases originates at time $t=0$, at node $a$. We would like to know the time at which the first infectious individual arrives in location $b$, that is to say the earliest time such that $I_b(t)>0$. We denote this epidemic arrival time (AT) as $\tau_a^b$. 
If no infected individuals ever arrives at $b$ we say that $\tau_a^b=\infty$.
The distribution of $\tau_a^b$ will depend on the starting location $a$, the target location $b$, and the network parameters $\gamma, T$, $\alpha$ and $\beta$. We wish to understand this dependence and do so by considering the survival function ${S}_a^b(t) = \mathbb{P}(\tau_a^b>t)$. 

Consider `Patient Zero,' initially placed in location $a$ at time $t=0$. Now consider the state of the system a short time later, at time $t=dt$. 
With overwhelming probability, nothing will have happened during this small time window, and the  survival times will still be governed by ${S}^{b}_a(t)$.
With probability $dt \gamma T_{a,k}$ patient zero will have moved to location $k$, and the survival time distribution will instead be governed by ${S}^{b}_k(t)$. In the case where $k=b$, survival is now impossible, hence ${S}^{b}_b(t)=0$. Patient zero may also infect someone with probability $dt \alpha_a$. In this case, the probability of survival until time $t$ is given by the probability that \emph{neither} traveler arrives at $b$. As both travelers are independent and identical, this probability is given by $[{S}^{b}_a(t)]^2$. Finally, it is possible that the initially infected patient zero may simply recover. This happens with probability $\beta_a$. In this case, no infected individual will ever reach $b$, and survival probability is $1$ for all time.
%These transitions are summarised in figure \ref{fig:FullDiplicateHit}.

The probability of surviving for $t$ units of time after $dt$ is thus a superposition of the possible survival functions for the states the system could transition to.
By definition the probability of surviving for $t$ units of time after time $dt$ is also precisely equal to ${S}^{b}_a(dt+t)$. Stated algebraically we thus have:

% \begin{figure}[p]
%     \centering
% \centerline{\includegraphics[width=1.15\textwidth]{FullDuplicatHittingTimeDistrib_noEquation.pdf}}
%     \caption{
%     A representative schematic of the possible states of the system after some small time $dt$ has passed. Our infected traveller may (from top to bottom) travel to the target node, travel to some other node, remain stationary, infect a second person, or recover from illness. We thus find that, $S_a^b(dt+t)$ is a superposition of $S_b^b(t)=0$, $S_k^b(t)$, $S_a^b(t)$, $[S_a^b(t)]^2$ and $1$. This relation is given in algebraic form in equation \ref{eq:preMAIN}.
%     \label{fig:FullDiplicateHit}}
% \end{figure}

\begin{align}
    {S}^{b}_a(dt+t) \approx & (1- dt (\alpha_a + \beta_a + \gamma \sum_{k\neq a} T_{a,k}) ) {S}^{b}_a(t) \nonumber \\
        & + \gamma \sum_{k\neq a} dt T_{a,k}{S}^{b}_k(t) + \alpha_a dt [{S}^{b}_k(t)]^2 + \beta_a dt,
    \label{eq:preMAIN}
        \\
        \intertext{rearranging and taking limits we find,}
    \lim_{dt \rightarrow 0} \frac{ {S}^{b}_a(dt+t) - {S}^{b}_a(t)}{dt} ={}& \gamma \sum_{k \neq a} T_{a,k} \left[{S}^{b}_k(t)-{S}^{b}_a(t) \right] \nonumber
    \\
    & \alpha_a \left[[{S}^{b}_a(t)]^2-{S}^{b}_a(t) \right]
    +\beta_a \left[1-{S}^{b}_a(t) \right].\\
   \intertext{remembering that $T_{a,a} S_a^b(t)=-\sum_{b \neq a} T_{a,k} S_a^b(t)$, we can simplify to} 
    \dot {S}^{b}_a(t) =& \gamma \sum_k  T_{a,k} {S}^{b}_k(t) + \left(\beta_a-\alpha_a {S}^{b}_a(t) \right)\left(1- {S}^{b}_a(t) \right), \nonumber\\
    {S}^{b}_b(t) =&0 , ~~~~~{S}^{b}_{a\neq b}(0) =1.
    \label{eq:MAIN}
\end{align}
We refer to arrival times calculated using this method as `branching process arrival times' (BP AT). These equations are equivalent to those given by Goldie \& Coldman \cite{goldie_mathematic_1979}, who model the arrival of treatment resistant cancer cells via rare mutation (although in Goldie's case, a much smaller collection of `types' of individual are considered).

It is important to note that eq. (\ref{eq:MAIN}) {\it does not} model the internal state of the system. While normal differential equations might be constructed by modelling the internal `state' of the system at time $t$, and using this to model the system at $t+dt$, instead we merely observe that the survival function $S_a^b(dt+t)$ (whatever it might be), can be written as a superposition of the (as yet unknown) survival functions $S_k^b(t)$, and hence derive eq. (\ref{eq:preMAIN}). %Rather than construct a bridge one brick at a time (while standing on the bridge itself), we instead construct the entire span of the bridge, and slide it across the river while standing on the riverbank ($t=0$)
In so doing, we have lent heavily upon the assumption that system parameters are constant in time; that the behavior and survival curve associated with a single traveller observed at time $t$ is identical to the behavior and survival curve of a traveller observed at time $0$. 
%In our construction of eq. \ref{eq:MAIN} we have lent heavily upon the assumption that the behavior of the system is identical at all points in time.
For a discussion of how a similar approach might be taken in the context of time varying parameters, see Appendix \ref{app:TimeVary}.

\subsection{Example Networks}

In order to test eq. (\ref{eq:MAIN}), and get something of an intuitive handle on the behavior of our system, let us consider three example networks of successively increasing complexity. For each network we solve eq. (\ref{eq:MAIN}) numerically using Matlab's ode45 \cite{shampine_matlab_1997} for each target node, and compare to survival times observed in exact agent based simulations, where we track individual infection, migration and recovery events precisely. Where possible simulations are conducted using Gillespie's exact algorithm \cite{gillespie_exact_1977}. In cases where such an approach is computationally infeasible, we make use of $\tau$-leaping \cite{gillespie_approximate_2001}.

\begin{figure}[p]
    \centering
    \includegraphics[width=0.85\textwidth]{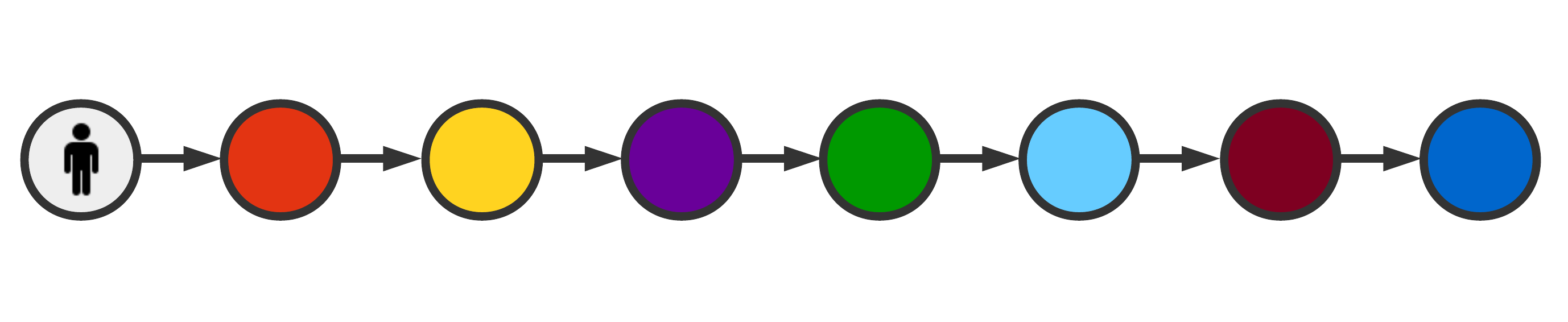}
    \includegraphics[width=\textwidth]{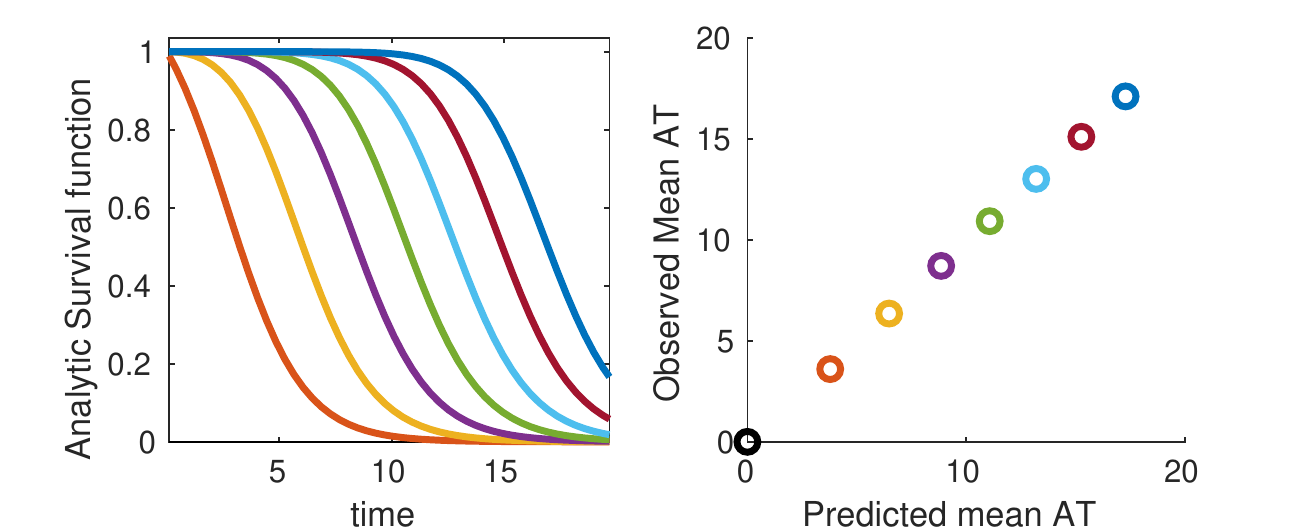}
    \caption{Here we consider the arrival time for an epidemic propagating down a simple chain (Top). (Left) Survival curves, calculated using eq. (\ref{eq:MAIN}) are approximately logistic, with the exception of the first step. (Right) Mean arrival times as predicted via eq. (\ref{eq:MAIN}) perfectly match mean arrival times as observed over the course of 5000 Gillespie simulations. Parameter values $\alpha=0.5$, $\gamma=0.1$. Nodes and survival curve are colour coded so as to match the network diagram, with the site of initial infection marked in black.}
    \label{fig:chain}
    \centering
    \includegraphics[width=0.39\textwidth]{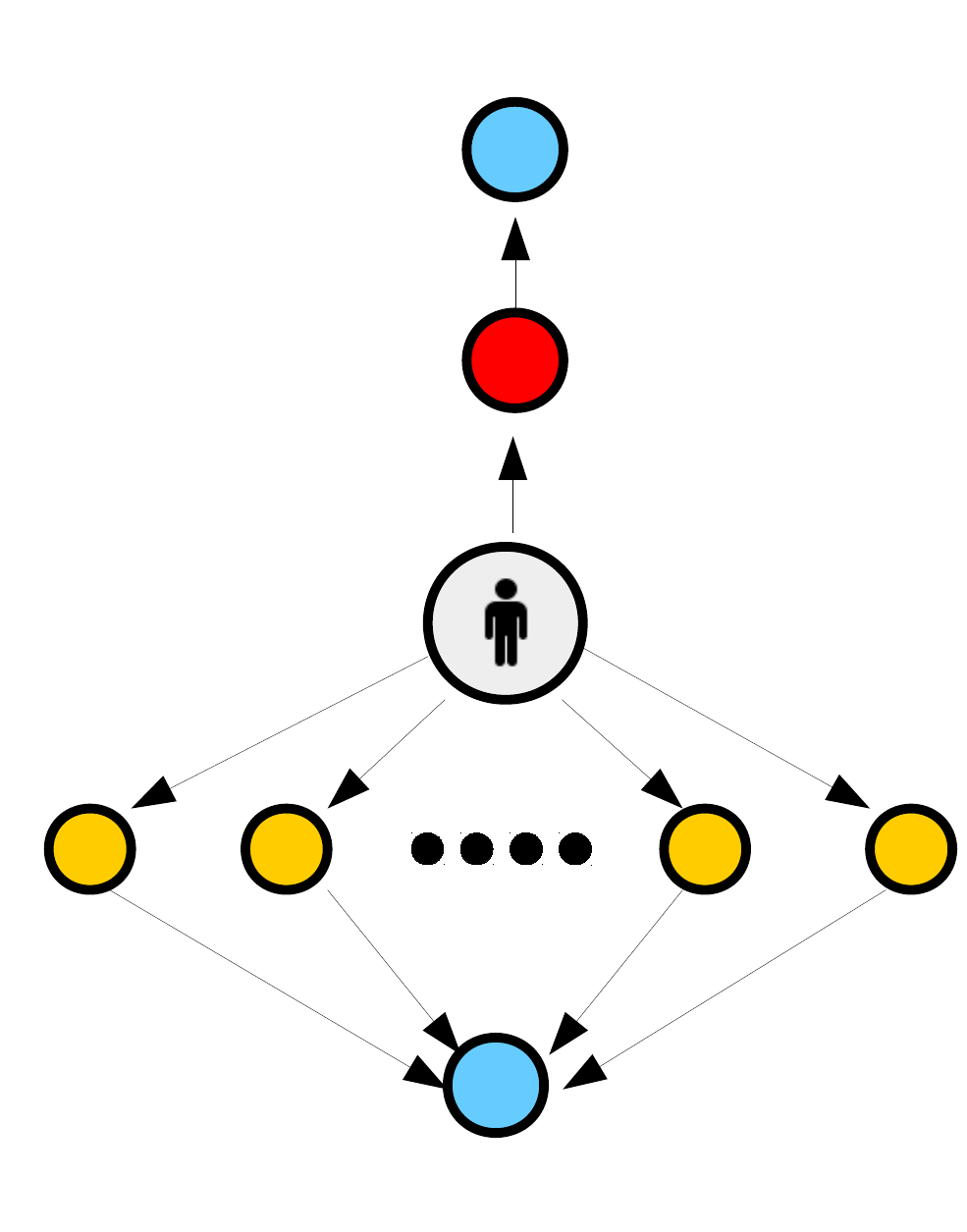}
    \includegraphics[width=0.59\textwidth]{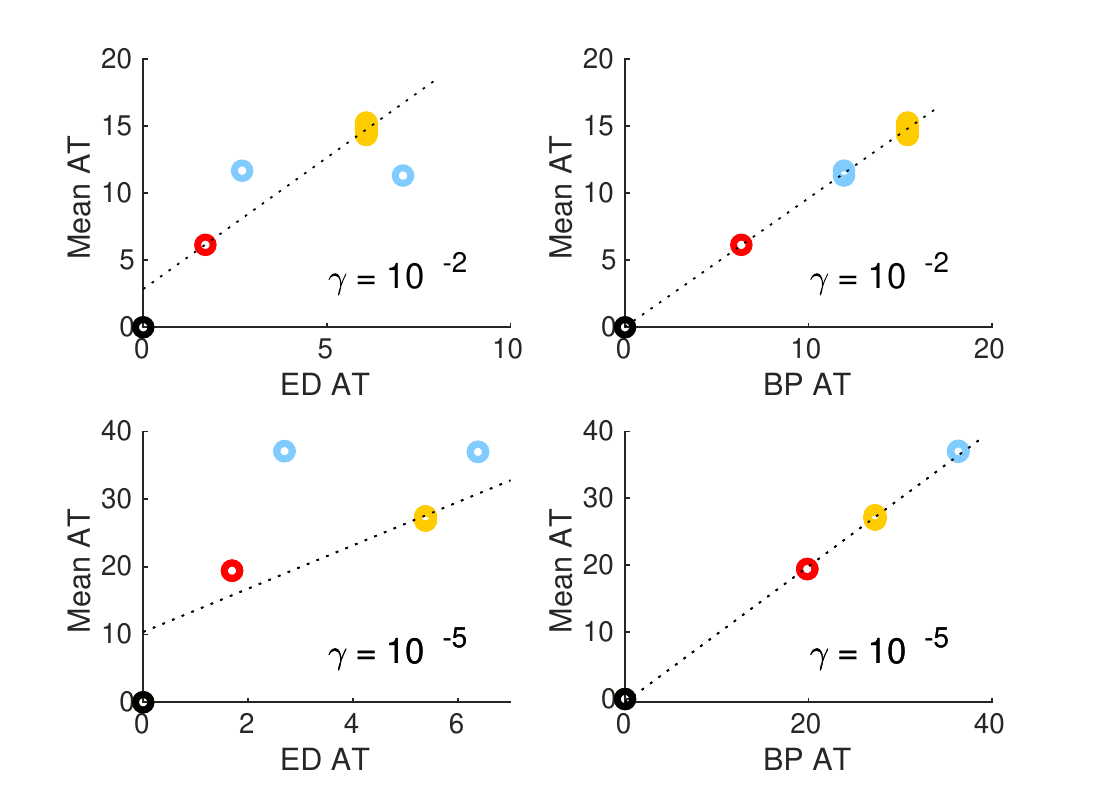}    \caption{(Left) The spinning top graph consists of $k+4$ nodes; a central starting node (grey) with no incoming edges, the initial site of infection, a `top' and `bottom' node with no outgoing edges (blue), a single ``up'' node directly between the center node and the top (red), and $k$ ``down'' nodes between the center node and the bottom (yellow). Infected individuals in the central node will travel either up or down at a rate of $\gamma$ - those travelling down select amongst the $k$ `down' nodes uniformly at random.
(Right)  Graphs of mean AT (averaged over 150 simulations) vs either effective distance (ED), or branching process arrival time (BP AT) for the Spinning top graph. In all cases, we assume $\alpha =0.5$, $\beta= 0$, $k=80$, and $\gamma$ as specified on the scatter plot.
Comparison of ED and BP AT demonstrates that ED distinguishes between the `top' and `bottom' node of the graph, while BP correctly handles multiplicity of paths, and assigns the pair identical ATs. We also see that BP AT is responsive to changes in $\gamma$, and correctly predicts the change in expected order of arrival when $\gamma$ is changed from $10^{-2}$ to $10^{-5}$. Simulations for this figure are carried out via $\tau$-leaping.
    \label{fig:spinTop} }
\end{figure}

As our first example, consider a simple chain graph, in which infection begins at one end of the chain, and is allowed to propagate from one node to the next (fig. \ref{fig:chain}). Here we observe that arrival time scales with number of steps taken, as might be expected. Survival curves are sigmoidal, which, as we will later see, is a typical behavior for reasonable parameter values. Mean arrival times as predicted by survival curves and as observed over 5000 Gillespie simulations are found to agree.

We next consider a `spinning top' graph (fig. \ref{fig:spinTop}); this graph is constructed in order to illustrate the effects of multiplicity of paths. Here one node of the graph is accessible through only a single possible path, while another node (at equal distance) can be accessed via a great multiplicity of paths, each with a correspondingly reduced probability.

Here we observe that the BP AT gives accurate predictions of arrival time, correctly predicting that the arrival time in the two most distant nodes is equal, regardless of the multiplicity of paths. Counter-intuitively, it is also observed that when $k$ (the number of paths) is greater than $1/\gamma$ , the mean arrival time in the outer blue nodes \emph{precedes} the mean arrival time in the intermediate yellow nodes. This occurs because, while the epidemic must reach \emph{some} intermediate node before passing to the outer blue node, it is also perfectly possible for a large number of intermediate nodes to be uninfected at this stage, resulting in a higher \emph{mean} arrival time. In the case $\gamma =10^{-2}$ this is precisely what occurs, and the ordering of mean arrival times inverts. The qualitative behavior of the system is thus sensitive to transport rate $\gamma$. All of this is correctly described and predicted by the BP AT; heuristic models that account only for a single shortest path (such as eq. (\ref{Eq:ED}) ) do not capture this effect (see fig. \ref{fig:spinTop}).

\begin{figure}[p]
    \centering
    \includegraphics[width=0.95\textwidth]{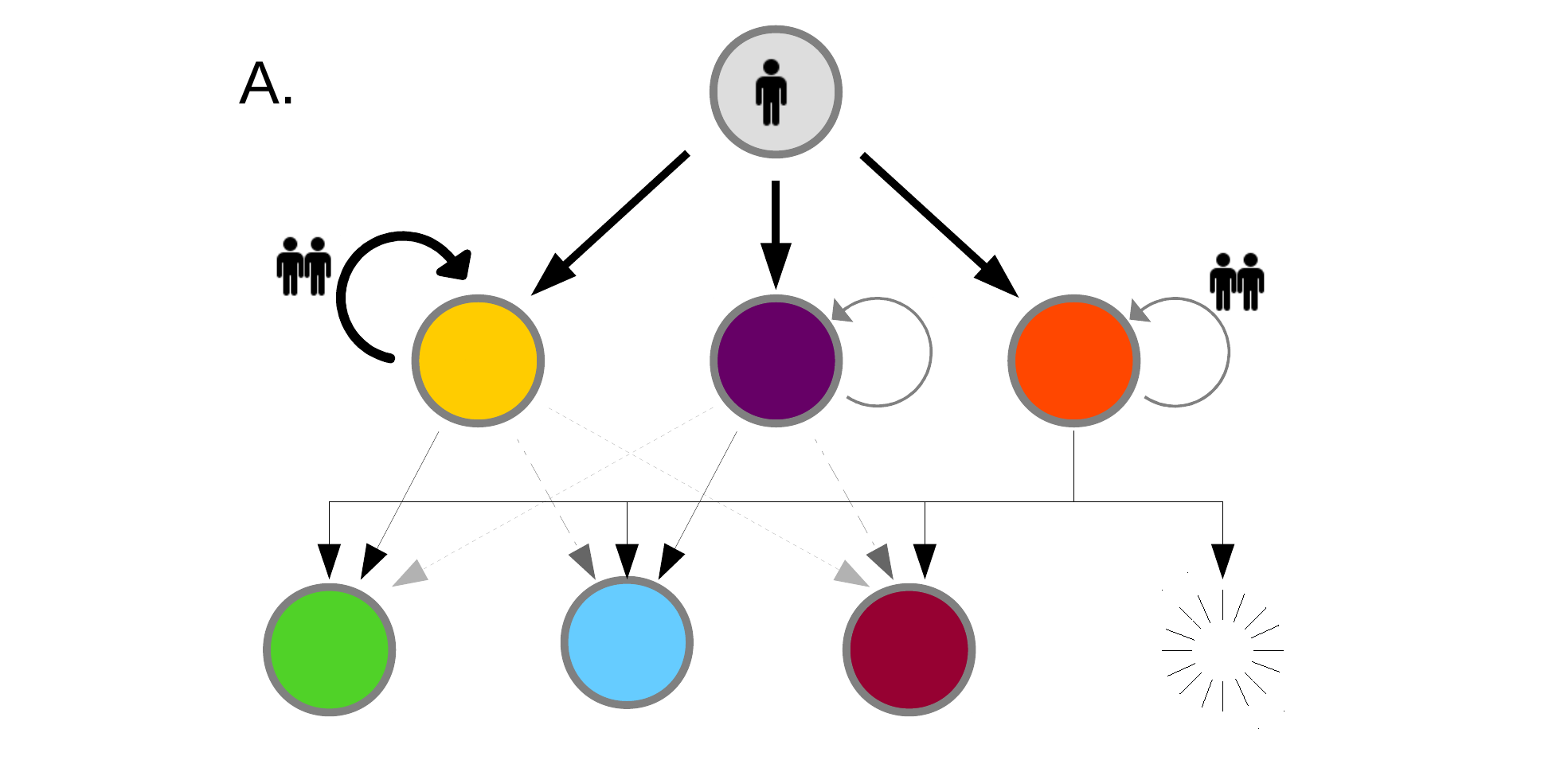}
    \includegraphics[width=0.95\textwidth]{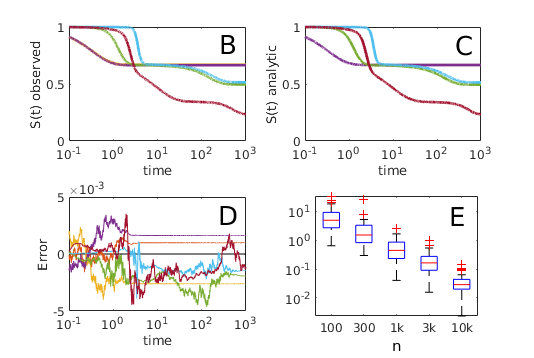}
    \caption{(A) Here we construct a complex network, designed to highlight the potential complexity of the survival curve $S(t)$. Patient zero is placed at the top of the graph and can transition to one of three `intermediate' nodes. These intermediate nodes have either high infection rate $\alpha=4$ or low infection rate $\alpha \approx 0.01$, and transition to the three `final' nodes at a variety of different rates ($1,10^{-2}$ or $10^{-4}$). The third intermediate node permits recovery from infection (transition to the empty circle) at rate $0.01$. While this network is not reflective of real world transportation networks, it does provide a useful illustration of the possible complexity of survival curves, and serves to demonstrate just how closely theory matches simulation even in these complex situations.    
    (B) Survival curves based on 25000 Gillespie simulations of the epidemic spreading process. Lines are colour coded to match the nodes in the network. (C) We solve equation \ref{eq:MAIN} and display the analytic survival curve ${S}^{b}_a(t)$ for each possible $b$, keeping $a$ fixed.
    (D) We plot the difference between the observed and analytic survival curves over time. (E) We sample $n$ of our 25000 simulations, then calculate $\sum \int |\text{error}|^2 dt$ when comparing observed and predicted survival probability. Here we integrate over time, and sum over possible arrival locations.
    }
        \label{fig:PerfectCompare}
\end{figure}

As a third example, in order to illustrate the possible complexity of arrival time distributions, we construct an artificial network with a wide spread of transportation, infection and recovery rates, and compare numerical solutions of equation \ref{eq:MAIN} with survival curves observed over the course of 25000 Gillespie simulations\cite{gillespie_exact_1977}. In this case we observe that survival curves exhibit rich, complex behavior. Nonetheless, equation \ref{eq:MAIN} correctly predicts the distribution of arrival times as observed in direct agent based simulation (figure \ref{fig:PerfectCompare}) over multiple time scales.

The branching process approach gives similar accuracy over a wide variety of parameter values and network structures. A gallery of such comparisons is provided in Appendix \ref{app:gallery}. 

% As a brief sanity check on Eq. \ref{eq:MAIN}, we take a small random network ($N=50$), with variable $\alpha$ ($0.5 \pm 0.1$) and $\beta$ ($0.24 \pm 0.06$), and compare numerical solutions of equation \ref{eq:MAIN} with survival curves observed over the course of 15000 Gillespie simulations\cite{gillespie_exact_1977}; All simulations are run until either every location has been visited at least once, or the entire infected population goes extinct. These results are summarized in figures \ref{fig:PerfectCompare}. Similar accuracy is observed for all tested parameter values and all tested network structures. A gallery of such comparisons, along with a detailed description of our Simulation scheme is provided in Appendix \ref{app:gallery}. 

\section{Derivation of Mean and Variance in Arrival Time}
\label{sect:meanVar}
Equation \ref{eq:MAIN}, while accurate, is somewhat cumbersome and does not provide any closed form for values of interest such as  expected arrival time and variance in arrival time. 
If we place no constraints on $\alpha$, $\beta$ and $\gamma T$ then it is easy to construct survival curves with complex topology such as fig. \ref{fig:PerfectCompare}, and it can be shown that \emph{any} valid survival curve (that is any curve of the form $0\le f(t) \le 1, f'(t)\le0,f(0)=1$) can be well approximated, for suitably chosen network and parameter values (see appendix \ref{app:ArbitraryCurves}). 
While mathematically interesting, such results are of little use to epidemiologists. More useful results can be obtained by placing some mild constraints on parameter values based on real world understanding.

The first assumption that we make is that we are not interested in epidemics that go extinct before spreading; we care only about those epidemics that reach international prevalence. For simplicity in the following discussion, this is achieved by assuming $\beta_k=0$ for all $k$. A discussion of non-zero $\beta_k$ can be found in Appendix \ref{app:BetaNonZero}. %In what follows, we redefine $\alpha_k$ to be the \emph{net} infection rate associated with our epidemic (as opposed to the total infection rate).

The second assumption that we make in what follows is that epidemic dynamics (infection and recovery) take place on on the time scale of days and weeks, while international travel is a rare occurrence, the vast majority of the human population taking international flights at most once or twice per year\footnote{With some individuals travelling significantly more often, and many others never flying at all.}. Stated algebraically we have $\gamma T \ll \alpha$; travel is rarer than infection. This mimics the `rare mutation' assumption made by Goldie \& Coldman\cite{goldie_mathematic_1979}%\todo{I NEED to get my hands on this initial paper! So far I've seen references to it but can't ACCESS the damn thing} 
when studying the the development of treatment resistant cancer cells, and rules out slow epidemics such as HIV.

We further assume that infection rate is equal across all locations: $\alpha_k=\alpha$. This greatly simplifies notation, and has limited impact upon our final conclusions.

Taken together, these three assumptions ($\beta_k=0, \alpha_k=\alpha, \gamma T_{ij} \ll \alpha$ are sufficient to avoid the more complex survival curve dynamics, as observed in fig. \ref{fig:PerfectCompare}.
Instead, we find $\dot {S}^{b}_a(t) \approx -\alpha {S}^{b}_a(t) \left(1- {S}^{b}_a(t) \right)$ and the arrival time distribution is well approximated by a logistic function:
\begin{align}
    {S}^{b}_a(t) \approx &  \frac{1}{1 + e^{\alpha (t-\mu_a^b)}}+O(\gamma).
    \label{eq:SmallGamApprox}
\end{align}
Here $\mu_a^b$ is the mean arrival time (AT) for an epidemic starting in node $a$ and spreading to $b$. Determination of $\mu_a^b$ is of some considerable interest. This value is dependent upon the dynamics of the system when $(1- {S}^{b}_a(t)) = O(\gamma)$; that is to say the region where the transport $\gamma T S^b_a(t)$ is no longer overwhelmed by logistic decay in probability. 

\subsection{Mean arrival time}
Determination of $\mu_a^b$ can be seen as equivalent to the question of determining when the exponential term of eq. (\ref{eq:SmallGamApprox})  is equal to $1$. For the sake of convenience, we label this term ${Q}^{b}_a(t)$, and make use of the change of variables ${S}^{b}_a(t)=1/(1+{Q}^{b}_a(t))$. This leads to the differential equation:
\begin{align}
    \dot {Q}^{b}_a(t) \approx & \alpha {Q}_a^b(t) -[1+{Q}_a^b(t)]^2 \gamma \sum_k T_{a,k} \frac{1}{1+{Q}_k^b(t)}  
    \label{eq:FirstQ}
\end{align}
Given Eq. \ref{eq:SmallGamApprox}, the mean arrival time $\mu_a^b$ satisfies ${Q}^{b}_a(\mu_a^b)=e^{\alpha(\mu_a^b-\mu_a^b)}=1$. Because ${Q}^{b}_a(t)<1$ for $t<\mu_a$, and $\gamma T \ll \alpha$, we can safely assume $\alpha {Q}_a^b(t)\gg \gamma (2{Q}_a^b(t) +[{Q}_a^b(t)]^2)$ in our region of interest, hence allowing the approximation  $[1+{Q}_a^b(t)]^2 \approx 1$. Because $ \sum_j T_{ij} =0$, we can restate eq. (\ref{eq:FirstQ}) as:
\begin{align}
    \dot {Q}^{b}_a(t) \approx & \gamma \sum_k T_{a,k} \frac{{Q}_k^b(t)}{1+{Q}_k^b(t)}  + \alpha {Q}_a^b(t).
    \label{eq:Qequation}
\end{align}
At this stage it may be tempting to make the further approximation 
\begin{align}
 \dot {Q}^{b}_a(t) \approx&  (\gamma T + \alpha I) {Q}_k^b(t) \label{eq:MatrixExpQODE} \,,\\
 \intertext{using the matrix exponential $e^M=I+M+M^2/2!+...$ we solve to find}
  {Q}^{b}_a(t) \approx&  e^{(\gamma T + \alpha I)t} {Q}_a^b(0)\,.
  \label{eq:MatrixExpQ}
\end{align}
Here we observe that eq. (\ref{eq:MatrixExpQODE}) and (\ref{eq:MatrixExpQ}) bear a striking resemblance to the calculation of the expected number of infections at $b$, given an initial epidemic starting at $a$, as studied by Chen et al. \cite{chen_estimating_2018}. Under this interpretation of ${Q}_a^b(t)$ we are effectively approximating our arrival time as the time at which the {\it expected} number of infected individuals in our target node, $b$, reaches one. %While Chen et al. consider the exponential growth of our infected population, coupled to the flow of individuals forward through the network, here we instead end up studying the flow of ``arrival probability'' \emph{backwards} through the network from our target location. 

\begin{figure}[h]
    \centering
\centerline{    \includegraphics[width=1.45\columnwidth]{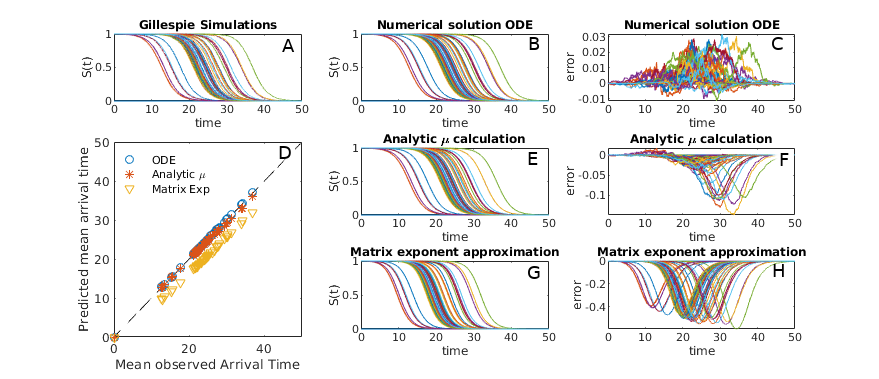}}
    \caption{Comparison of arrival times arrival times as observed in 5000 direct Gillespie simulations of epidemic spread (A) to survival times as predicted using either: numerical solutions Eq. \ref{eq:MAIN} (panel B),  logistic curves centered according to $\mu$ Eq. \ref{eq:ComplexQDefinition}  (panel E), or logistic curves with $Q(t)$ as defined by simple matrix exponentiation (Eq.\ref{eq:MatrixExpQ}, panel G). In the right hand column we give the difference between predicted and observed survival probabilities for each method. (D) Comparison of mean AT as observed in 5000 simulations with the mean AT as predicted by the three numeric methods shown in the adjoining panes.}
    \label{fig:BigPanelFigure}
\end{figure}

In cases where a single path of length $d$ dominates the spread of our epidemic from $a$ to $b$ we can make the further approximation:
\begin{align}
 {Q}_a^b(t)= 1= & e^{\alpha t} \left[I + \frac{\gamma T t}{1!}+\frac{\gamma^2 T^2 t^2}{2!}+... \right]_{a,b} \approx e^{\alpha t} [T^d]_{a,b} \frac{\gamma^d t^d}{d!}\\
  \alpha t \approx&  -\log \left( \frac{\gamma^d T^d_{a,b} t^d}{d!} \right).\\
  \intertext{Choosing to ignore a number of small terms, we find}
  \approx& -d \log{\gamma} - \log \left(\prod T_{i,j} \right) = - \sum \log{\gamma} + \log  T_{i,j},
  \label{eq:shortPath}
  \end{align}
where here $T_{i,j}$ are the transport rates along each link of the most probable path epidemic path. We thus reconstruct the results of shortest path methodology similar to Gautreau's original study \cite{gautreau_arrival_2007} and the work of Brockmann \& Helbing \cite{brockmann_hidden_2013}. 
More detailed discussion of this matrix exponential approach and its relation to previous work is provided in Appendix \ref{app:chen}. Survival curves as predicted by equation \ref{eq:MatrixExpQ} are given in figure \ref{fig:BigPanelFigure}. 

%Chen et al. solve the transcendental Eq. \ref{eq:MatrixExpQ} by expanding out the matrix exponential as a power series and considering only the dominant terms; an alternative approach (which we make use of) is to calculate the eigenvalue decomposition of $\gamma T + \alpha I$ and use Newton's method to solve $e^{(\gamma T + \alpha I)t} {Q}_a^b(0)=1$ directly (see figure \ref{fig:BigPanelFigure}).

While the approximation $Q/(1+Q) \approx Q$ is convenient in that it allows us to linearize the equation, more accurate results can be obtained by retaining $1/(1+Q)$ and following on from equation \ref{eq:Qequation} directly. Applying integrating factors we find,
\begin{align}
    e^{-\alpha t} {Q}^{b}_a(t) \approx & \int_0^t e^{-\alpha \tau}
 \gamma \sum_k T_{a,k} \frac{{Q}_k^b(\tau)}{1+{Q}_k^b(\tau)} d\tau.\\
 \intertext{Substituting the ansatz $Q_k^b(\tau) = e^{\alpha (\tau-\mu_k^b)}$ for $k\neq a,b$, along with $\frac{Q_b^b(t)}{1+Q_b^b(t)}= 1-S_b^b(t) = 1$, we find}
    e^{-\alpha t}{Q}^{b}_a(t) \approx&  \gamma T_{a,b} \int_0^t e^{-\alpha \tau} d\tau  + \sum_{k\neq b} \gamma T_{a,k} \int_0^t e^{-\alpha \tau} \frac{e^{\alpha(\tau-\mu_k^b)}}{1+e^{\alpha(\tau-\mu_k^b)}} d\tau.
    \label{eq:Convolution}
    %{Q}^{b}_a(t) \approx & \gamma T_{a,b} \int_0^t e^{\alpha(t-\tau)} d\tau+    \sum_{k\neq b} \gamma T_{a,k} e^{\alpha (t- \mu_k^b)} \int_0^t \frac{e^{-\alpha(\tau-\mu_k^b)}}{e^{-\alpha(\tau-\mu_k^b)}+1} d\tau\\
    \intertext{Multiplying through by $e^{\alpha t}$ and integrating provides}
        {Q}^{b}_a(t) \approx & \gamma T_{a,b} \frac{e^{\alpha t}-1}{\alpha} +    \sum_{k\neq b} \gamma T_{a,k} \frac{e^{\alpha (t- \mu_k^b)}}{-\alpha}  \text{log}\left(\frac{e^{-\alpha t}+e^{-\alpha \mu_k^b}}{1+e^{-\alpha \mu_k^b}}\right).
        \label{eq:ComplexQDefinition}
\end{align}

Our approach in this equation is to explicitly solve for $Q_a^b(t)$ while treating all other survival curves as logistic.
Because $e^{\alpha(\mu_a^b-\mu_k^b)}$ will be non-negligible only when $\mu_k^b<\mu_a^b$, it is possible to calculate arrival time from smallest $\mu_k^b$ up to largest, using knowledge of the previously derived $\mu$ to inform calculation of later $\mu$. In many cases eq. (\ref{eq:ComplexQDefinition}) is dominated by a single term, allowing us to make the approximation ${Q}^{b}_a(\mu_a^b)=1 \approx  \gamma T_{a,k} e^{\alpha (\mu_a^b- \mu_k^B)} \mu_k^b$ for some $k$. Taking logs of both sides we find $(\mu_a^b- \mu_k^b) = -\log(\gamma T_{a,k}\mu_k^b )/\alpha \approx- [\log(\gamma)+\log(T_{a,k})]/\alpha$. Summing over multiple such steps reconstructs a shortest path type metric, similar to eq. (\ref{eq:shortPath}). 
Even in cases where shortest path approximations do not apply, Newton's method can be easily applied to solve ${Q}^{b}_a(\mu_a^b)=1$, allowing us to determine all $\mu_a^b$ for any given $b$. 

Eq. \ref{eq:ComplexQDefinition} can thus be used to accurately approximate $Q_a^b(t)$ and $S_a^b(t)$ without resorting to computationally expensive numerical ODE solvers; this is particularly helpful in regions of parameter space where high sensitivity near $t=0$ render more direct numerical methods unstable.

%In many cases eq. \ref{eq:ComplexQDefinition} is dominated by a single term, allowing us to make the approximation ${Q}^{b}_a(\mu_a^b)=1 \approx  \gamma T_{a,k} e^{\alpha (\mu_a^b- \mu_k^B)} \mu_k^b$ for some $k$. Taking logs of both sides we find $(\mu_a^b- \mu_k^b) = -\log(\gamma T_{a,k}\mu_k^b )/\alpha \approx- [\log(\gamma)+\log(T_{a,k})]/\alpha$. Summing over multiple such steps once again reconstruct the shortest path arrival time $\alpha \mu_a= \min \sum [\log(\gamma)+\log(T_{i,j})]$, where here we take the minimum along all possible paths from $a$ to $b$, and summation is conducted over the steps of a given path.%; that is to say, in cases where eq. \ref{eq:ComplexQDefinition} is dominated by a single term, we recreate a shortest path approach similar to those given by both Gautreau et al \cite{gautreau_arrival_2007} and Brockmann \& Helbing \cite{brockmann_hidden_2013}. The `distance' between adjacent nodes is dependent on network geometry and is given by $\log(\gamma)+\log(T_{a,k})$, while the `velocity' of travel depends on the disease of interest, and is given by $\alpha$.

A comparison of the various approaches described in this section is given in Fig. \ref{fig:BigPanelFigure}. Numeric solutions of equation \ref{eq:MAIN} give the best results (when using high accuracy ODE solvers), while predictions based on simple matrix exponentiation (Eq. \ref{eq:MatrixExpQ}) are the least accurate. All three approaches are highly correlated with mean arrival times as observed in full agent based simulations however.  When we are interested in relative rather than absolute arrival time, any of the three approaches will suffice, and matrix exponential based approaches are the fastest (see appendix \ref{app:chen} for details on efficient computation). When more accuracy, or knowledge of full probability distributions, is required, we recommend using either equation \ref{eq:MAIN} or \ref{eq:ComplexQDefinition}.

\subsection{Variance}
\label{sect:var}
Another value of practical interest is the variability in arrival time; recognising the difference between $52 \pm 3$ days and $52 \pm 30 $ days allows us to understand how \emph{precise} we expect arrival predictions to be.  When $\gamma T \ll \alpha$, we can use standard logistic distribution results \cite{balakrishnan_handbook_2013} to determine    $\text{var}(\tau_a^b) =\frac{\pi^2 s^2}{3}= \frac{\pi^2}{3\alpha^2}$, where $s$ is the `scale parameter' of the logistic distribution, in our case equal to $1/\alpha$. In plain language, we find that variance is high for slowly growing epidemics, and low for fast growing epidemics, where the infected population spends shorter periods of time at intermediate levels.

This determination of variance leads naturally to the question of co-variance; if our epidemic arrives three days earlier than expected in Paris, will it also inevitably arrives three days early in Istanbul or S\~{a}o Paulo? Or are the two arrival times relatively independent? This question is of particular importance because, generally speaking, we do not \emph{know} when an epidemic has started, and hence can only ever compare arrival times in different cities relative to one another.

Unfortunately, by its nature ${S}^{b}_a(t)$ contains minimal information about the state of the system; calculation of such state information using probability generating function techniques \cite{miller_primer_2018} would require us to take infinitely many derivatives of ${S}^{b}_a(t)$ with respect to our initial conditions ${S}^{b}_a(0)$. It would appear that full co-variance and correlation information is thus inaccessible.

One avenue that we \emph{can} take is to examine the variance in arrival time given a larger starting population - for example $p=50$, with all individuals starting at location $a$. Such an assumption effectively removes the large levels of variance associated with the epidemics ``initial take off'' time (the time taken to grow from $p=1$ to $p=50$), and leaves only the variation associated with the spreading process throughout our network (and subsequent take off time in the various locations the epidemic emigrates to). Given that we typically do not get to observe the true epidemic start data in practice, such an assumption is likely to better reflect real world data.

The survival time distribution for a starting population of $p$ is simply $ [S_a^b(t)]^p$. It can be shown (see appendix \ref{app:varianceHighPop}) that for an initial population $p$, the arrival time has variance 
\begin{align}
\text{var}(\tau)&=  \frac{\pi^2}{3 \alpha^2} - \sum_{k=1}^{p-1} \frac{1}{\alpha^2 k^2}
 \label{eq:varianceVsPop}
\end{align}
In the limit $\sum_{k=1}^{\infty} (\alpha k)^{-2}=\pi^2/6 \alpha^2$.

\begin{figure}[h]
    \centering
    \includegraphics[width=0.49\textwidth]{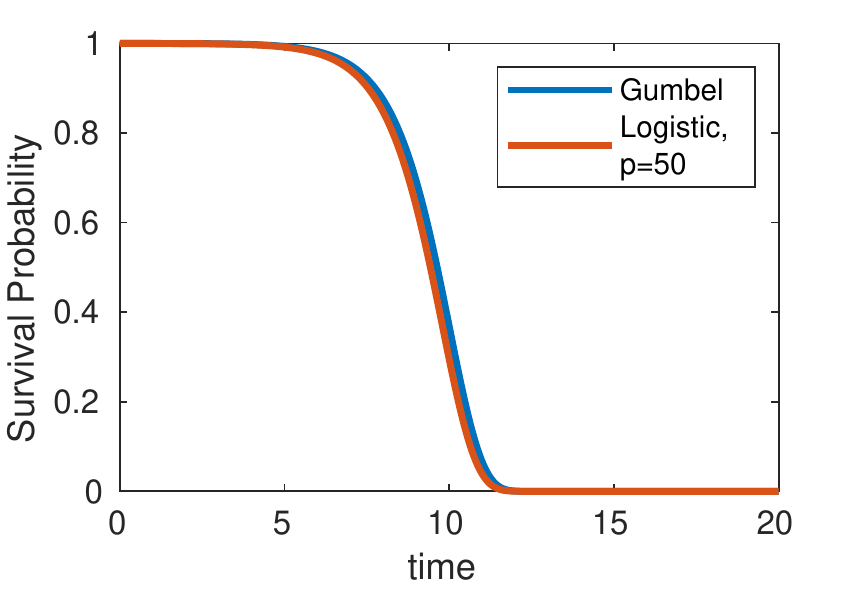}    \includegraphics[width=0.49\textwidth]{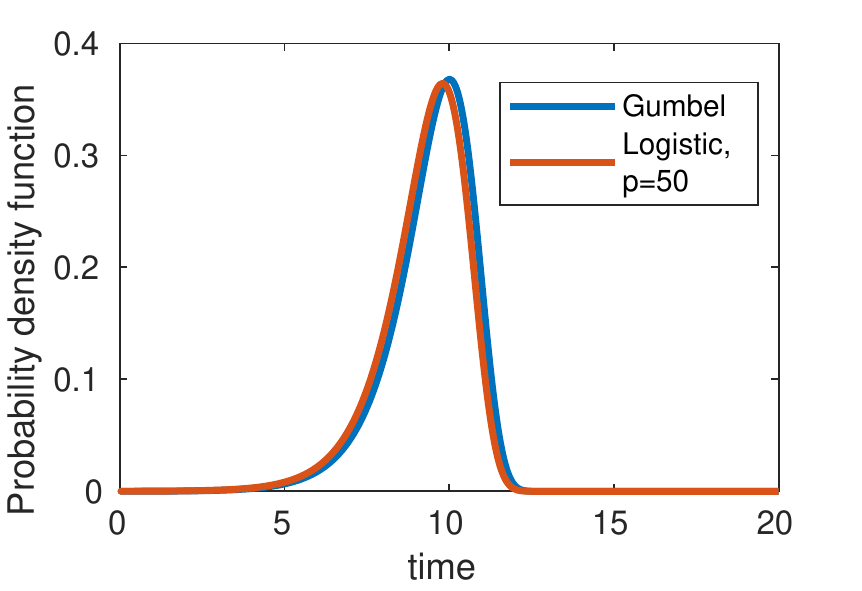}
    \caption{Here we compare the Gumbel distribution (the survival curve identified by Gautreau et al. when modelling population as continuous \cite{gautreau_global_2008}), and the exact survival curve assuming a initial population $p=50$, as discussed in section \ref{sect:var}. (Left) Survival curve for Gumbel distribution ($\exp[-e^{(t-\mu)}] $) or logistic like arrival times with $p=50$ ($(1+e^{(t-\mu)})^{-p} $) (Right) Probability density function for the same. Here we compare only shape, hence, in this example the curves have different $\mu$ parameter; this acts only to slide the mean value so as to make the curves more comparable.}
    \label{fig:GumbelVsLogistic}
\end{figure}

Of further note, we also observe that as $p$ increases the $p^{th}$ power of the logistic curve ($(1+e^{(t-\mu)})^{-p} $) approaches the Gumbel distribution (figure \ref{fig:GumbelVsLogistic}) as observed by Gautreau et al. \cite{gautreau_arrival_2007} in their original study of the epidemic arrival time process. The Gumbel distribution can be loosely thought of as an exponential waiting time with exponentially increasing rate parameter (corresponding to the growing population), and has cumulative distribution function of the form $\exp[e^{-(t-\mu)\beta}]$. The continuous population assumption implicit in this result is valid when $p\ge 100$, but causes noticeable discrepancies for $p \le 5$.

\section{Robustness and Limitation}
\label{sect:limits}
So far we have explored the branching process model in what may be considered close to optimal conditions; we have assumed that $T$, $\alpha$ and $\gamma$ are exactly known, and that the susceptible population of each node is large enough so as not to limit epidemic growth and spread. Each of the above assumptions may be violated in one context or another, and for this reason it is critical to understand how far each can be stretched. Such understanding sheds light not only on the BP AT itself, but also on past distance metrics, which we have shown to rely on a similar theoretical foundations.

\begin{figure}[p]
    \centering
    \centerline{ \includegraphics[width=1.05\columnwidth]{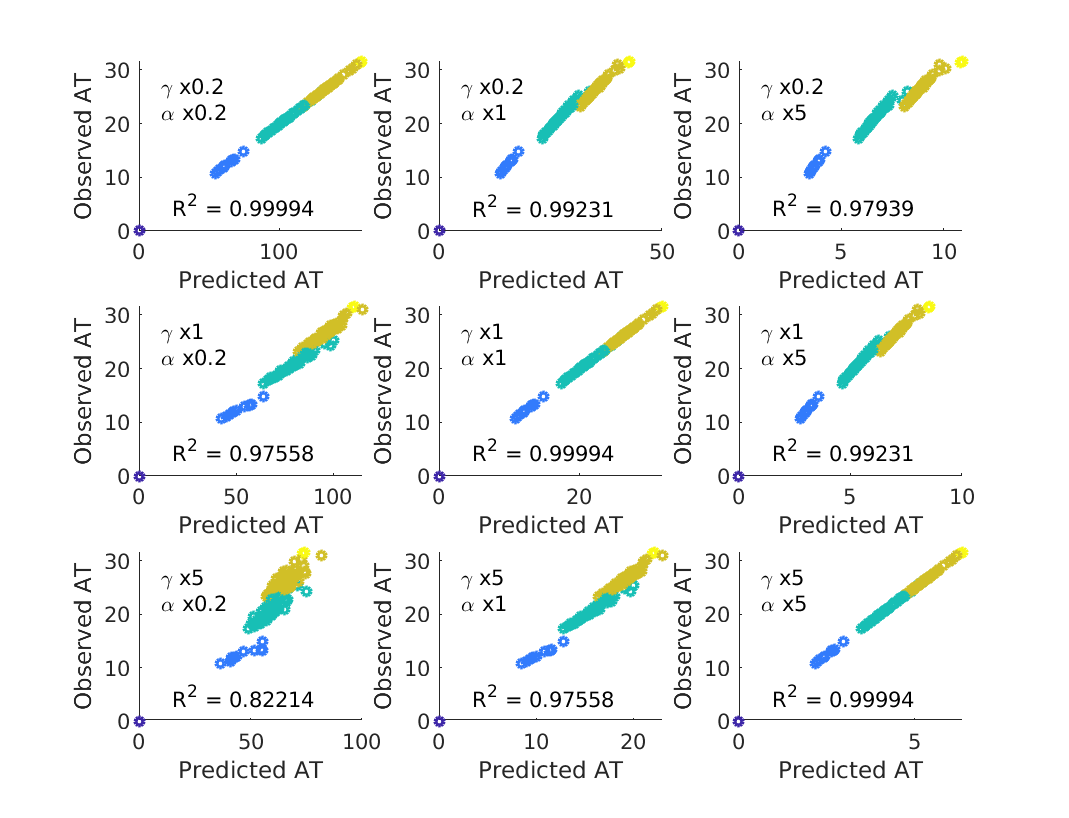}}
    \caption{Calculating the BP AT using incorrect growth and transport parameters leads to suboptimal results. Here we compare the Predicted mean AT (according to eq. (\ref{eq:MAIN})) with mean observed AT (averaged over 5000 agent based Gillespie simulations) in the case of using $\alpha$ and $\gamma$ which are either a factor of five too high or to low compared to their true values ($\alpha=0.5, \gamma=10^{-3}$). Data is coloured according to the minimum number of flights from origin to target. While errors change the magnitude of predicted AT (note varying axis scales), the effect on \emph{relative} AT predictions is minor in eight out of the nine cases considered.}
    \label{fig:wrongGammaAlpha}
\end{figure}

The first, and most likely assumption to be violated in practice is the notion that we know $\alpha$ and $\gamma$. While $\gamma$ can be determined to some reasonable level of accuracy via commercial flight data, $\alpha$ is a number that will vary from illness to illness and may be known only to a limited degree of accuracy, particularly in the early stages of a pandemic. Fortunately the model turns out to be robust against variations in both $\alpha$ and $\gamma$ -- even when varying these parameters by a factor of five compared to the `true' values used in simulations, predicted ATs remain highly correlated with those observed in simulation (see figure \ref{fig:wrongGammaAlpha}). When using incorrect $\alpha$ the constant of proportionality between $\tau_i$ and observed ATs is no longer equal to one, suggesting that the BP model robustly estimates $\alpha \tau_i$, such that the \emph{relative} time of arrival at any two locations is largely independent of $\alpha$ and $\gamma$, but that incorrect estimates of $\alpha$ will lead to corresponding inaccuracy in the \emph{absolute} values of $\tau_i$; if $\alpha$ is estimated a factor of two too high, then $\tau_i$ will be a factor of two too low.

\begin{figure}[p]
    \centering
    \centerline{ \includegraphics[width=1.15\columnwidth]{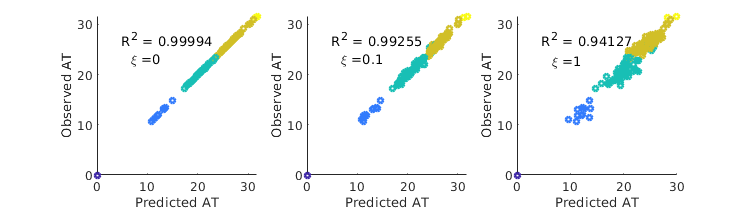}}
    \caption{Comparing the results of Gillespie simulations with the predictions made using a perturbed version of the original random network. We are still able to accurately predict AT, even in the case of significant perturbations to $T$, indicating that BP AT is robust even when link weights are increased or decreased by an order of magnitude. As previously, nodes are colour coded based on the number of steps to the initial infection site. Note that here, after noise is applied to calculate $\hat T$, we demand symmetry, and recalculate diagonal entries so as to preserve mass. Similar results are observed for scale-free networks.}
    \label{fig:ErrorF}
    \centering
    \centerline{
    \includegraphics[width=0.65\columnwidth]{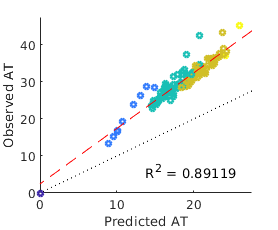} }
    \caption{When populations are very small, the exponential growth assumption that the BP AT relies upon no longer gives accurate predictions. Here we plot expected arrival time according to the BP model, against the average results of 5000 Gillespie simulations on a random graph of 135 nodes, where the susceptible population in each node is $18.2\pm7.5$. We assume infection rate $\alpha= 0.5$, and recovery rate $\beta=0$ (for small populations, $\beta>0$ generally leads to extinction). 
    The dotted black line gives the `1:1' line that would be expected of accurate predictions. The dashed red line gives the line of best fit, as used to calculate $R^2$.}
    \label{fig:smallPopBreakdown}
\end{figure}

Another systematic source of error is inaccuracies in flight network data. These can be produced due to uncounted or unregistered flights (or alternative forms of transportation), out of date data, or as a direct result of changes to individual flight plans as a result of the epidemic itself. In order to test robustness to noise in available flight data, we compare the results of full Gillespie simulation using flight network $T$, to ATs predicted using the perturbed matrix $\hat F$, where $\hat T_{ij}= T_{ij} e^{ N(0,\xi) } $. Here $\xi$ is a constant representing noise level. As can be seen in Fig.  \ref{fig:ErrorF}, the resulting errors are modest. 

While we have thus far considered arrival times in the context of the global aviation network, effective distance measures have also been considered on a much smaller scale in order to study the spread of antibiotic resistance between wards within a hospital \cite{friso_coerts_network_2018}. In this case, assuming unbound growth becomes problematic, as the susceptible population in each node is rather small; a single ward might contain (for example) $10-30$ beds, a far cry from the tens or hundreds of thousands found when each node represents an entire city.
When the local susceptible population is small, the epidemic is frequently no longer in an exponential growth phase by the time it spreads to a neighbouring node; `branching random walks' are no longer independent because it is impossible for person $A$ to infect someone who has already been infected by person $B$. Population saturation inevitably leads to less accurate predictions, see figure \ref{fig:smallPopBreakdown}. 

In practice, for the SIR type models the parameter window where this concern is relevant is relatively narrow; infections which grow quickly enough to violate our assumptions are liable to burn through the entire local population and drive themselves to extinction before spreading between nodes. For SIS and SI type models, or any infection which is expected to reach a local endemic equilibrium, it is important to determine whether exponential growth is a good approximation before making use of the BP AT, or any distance metric that relies upon the same underlying unbound growth assumption.

\section{Real World Data}
\label{sect:world}
\begin{figure}[p]
    \centering
\centerline{    \includegraphics[width=1.10\columnwidth]{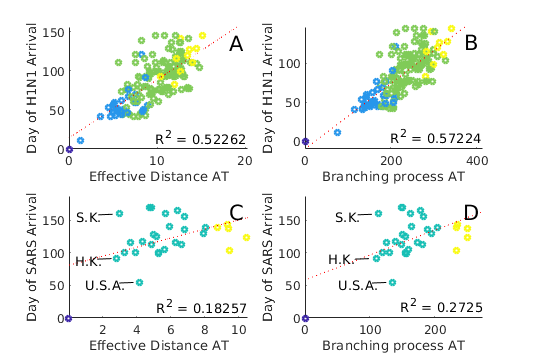}}
    \caption{Comparison of distance metrics to arrival times in real world epidemics for the 2009 H1N1 epidemic (top) and the 2003 SARS epidemic (bottom).
    (A) We attempt to reconstruct Figure 2D of Brockmann \& Helbing (2013) \cite{brockmann_hidden_2013}; we find that numerous details of the general \emph{shape} are reconstructed (for example the `branch' on the right hand side mid way up, and the cloud of seperated points to the left at the top), but that the figure as presented in the original paper is slanted significantly closer to the diagonal. (B) For suitably chosen $\alpha$ and $\gamma$ values, %TODO (Discuss)
    the BP method improves upon $R^2$, but is still limited by the accuracy of the underlying network, and the inherently noisy details of real world epidemiology. (C \& D) In case of the SARS epidemic we were not able to reproduce the extremely tight correlation previously reported \cite{brockmann_hidden_2013}. Once again, use of the BP metric yields improvements in $R^2$ compared to our reconstruction the ED metric. The three marked nodes in the bottom images corresponding to (from top to bottom) South Korea, Hong Kong, and the United States. As might be expected for distance measure based on flight data, both Korea and Hong Kong are `close' to China according to both metrics, while the USA is further away. This would appear to be in contrast to previously published results \cite{brockmann_hidden_2013}, which appear to indicate that the United States is closest to China, while Korea is rather distant. Data provided by Dirk Brockmann (private communication). 
}
    \label{fig:h1n1Compare}
    \label{fig:SarsCompare}
\end{figure}

Finally, while comparison to simulations provides an effective test bed, useful in terms of repeatability and certainty of data, we are, generally speaking more interested in the performance of models as they apply to the real world. Using flight data provided by Dirk Brockmann (private correspondence) we compare observed and predicted AT for both SARS and H1N1 outbreaks (see figure \ref{fig:h1n1Compare}). We consider predictions made by both the branching process method described here (Eq. \ref{eq:MAIN}) as well as Brockmann \& Helbing's effective distance metric (Eq. \ref{eq:OriBrock}, \cite{brockmann_hidden_2013}).

Ideally, in order to calculate the BP AT distribution, we would prefer to have access to the full transport matrix $T_{ij}$, the probability that an individual in location $i$ will travel to $j$ on any given day. Unfortunately such data is not available, and can not realistically be obtained. Instead, we have access to $F_{i,j}$; a matrix containing the total \emph{number} of passengers travelling from $i$ to $j$ during a given time period. In order to approximate $T_{i,j}$ we normalize $F$, hence calculating the probability, $P_i,j$, to fly from $i$ to $j$, \emph{conditioned} on the assumption that we board some plane in airport $i$, and multiply by a constant $\gamma$, representing the probability of boarding \emph{some} flight on a given day. In the real world, we might assume $\gamma$ to vary significantly from place to place, based on economic factors, prevalence of tourism, and the size of the population that any given airport is servicing. For the time being we approximate $\gamma$, and rely on the models low sensitivity to $\gamma$ to prevent difficulties.

We find that both BP AT and effective distance arrival time (`ED AT') methods give qualitatively similar results (see fig. \ref{fig:h1n1Compare}), and that these results would appear in many cases plausible: for example, in the case of SARS, we observe that predicted arrival times in both South Korea and Hong Kong are low, while the arrival time in the USA is predicted to be marginally higher. This result is entirely plausible given airline traffic between the respective countries. In practice, predictions for both BP AT and ED AT do not match observed arrival times: SARS is reported to have arrived in the USA $54$ days after the first reports in China (11/16/2002), and in South Korea a full $160$ after these first reports. For SARS, Arrival times as reported from real world data correlate at best weakly with predictions(note the small $R^2<0.3$ in fig. \ref{fig:h1n1Compare} C,D), indicating either inaccuracy in the data or alternatively some complexity in the real world process not accounted for in current models. $R^2$ values are somewhat better when predicting H1N1 ($R^2\approx 0.55$).

We find no method (including ED) is able to reproduce the very high $R^2$ values reported in Brockmann \& Helbing's original paper, and that we are unable to recreate their figures (Figures 2D and 2E in the original paper), instead observing a far broader scatter. It is unclear if the observed discrepancies is due to differences in implementation of the algorithm, differences in the underlying data, or some other cause. While it is possible that inaccuracies in WAN data are to blame for these discrepancies, this explanation seems unlikely; as demonstrated in figure \ref{fig:ErrorF}, arrival time predictions are generally stable, even to substantial changes in $F$. If we consider the effective distance metric (Eq \ref{eq:OriBrock}), increasing $D_{i,j}$ by $5$, as is needed to correctly predict SARS late arrival time in South Korea, would require a $P_{i,j}$ two orders of magnitude smaller than our current value. Such drastic changes appear implausible.
%More likely, it would see that the observed spread is the \emph{inevitable} result of the underlying stochastic process of the system.

%\section{SEIR model}

%\section{Time varying parameters}

\section{Conclusions}
Better understanding the spread of epidemics through the WAN allows for both real time forecasting (as has been used during the current COVID-19 pandemic), and also the possibility of network design, making changes to the WAN network so as to slow epidemic spread.

In studying the question of epidemic arrival time, a variety of models have been used, from the the most intricate agent based simulations\cite{broeck_gleamviz_2011} to the intuitively appealing `distance' based models \cite{brockmann_hidden_2013}, and a number of analytic approaches in between \cite{chen_estimating_2018, iannelli_effective_2017, gautreau_global_2008}.
While each of these models approaches the question of epidemic arrival time from a different view point, one common thread is the assumption of exponential growth; in the early stages of an epidemic, spread is governed by unbound epidemic growth, and rare transportation events. Population saturation, detailed viral dynamics, and a travellers tendency to return to their port of origin are ignored in all but the most detailed of models.

Given this basic premise, we were able to formulate the problem of epidemic arrival times in term of `Branching Processes' \cite{kimmel_branching_2015,goldie_mathematic_1979}, and calculate the full probability distribution of possible arrival times explicitly. These predictions match perfectly to corresponding simulations. 
If we further assume that air travel is rare and infection and recovery are common, it is possible to re-derive many of the results of past papers, and in some cases improve upon them. We are also able to give predictions in regions of parameter space where past methods are known to fail, and show that theoretical predictions provide a reasonable match to simulations, even when parameter values are altered significantly. 

Unfortunately, when comparing to real world data we observe that the predictive powers of ED metrics may be significantly lower than previously reported \cite{brockmann_hidden_2013}. When compared to real world date, ED and BP like methods predict roughly 50\% of variance in the cases of H1N1, and only 20\% of variance when compared to SARS-2003, significantly lower than would be predicted given the modest intrinsic variance of our models. While the Branching Process Arrival Time makes very few assumptions and gives exact results given the underlying model assumed, these findings are suggestive of a gap in our knowledge- either in the data available for epidemic arrival times or (more likely) in the model itself. It seems likely that future efforts would be best directed towards determining what real world factors are currently unaccounted for, and which of these are most critical.

%a disappointing, though not entirely unexpected result given the complex and contingent nature of global epidemic spread. Noise and chance are, as always, powerful forces within our world. 

%Unfortunately, when comparing to real world data we observe that the predictive powers of ED metrics may be significantly lower than previously reported \cite{brockmann_hidden_2013}. When compared to real world date, ED and BP like methods predict roughly 50\% of variance in the cases of H1N1, and only 20\% of variance when compared to SARS-2003, significantly lower than would be predicted given the intrinsic variance of our models. Given that the BP AT results studied here require no This would appear to be strong evidence that epidemic spread is governed by more complex processes than the simple

\section{Acknowledgements and Data Availability}
We gratefully acknowledge Dirk Brockmann for stimulating conversation, and for providing historical network and epidemic data.
Full code for simulation based figures is available on github \cite{jamieson-lane_alastair-jlepidemicspread_2020}. 
\appendix 

\section{Time Varying Parameters and flight networks}
\label{app:TimeVary}
Construction of equation \ref{eq:MAIN} relies heavily upon the assumption that the process is memoryless and does not keep track of time. 
%While normal ODEs are constructed by adding small slices of time, $dt$ on to the end of our function, $S(t)$ is instead constructed by adding slices of time to the \emph{beginning} of our function, and then sliding the entire function forward in time. The Markov assumption provides precisely the ``frictionless'' environment required for this sliding process to take place.

\begin{figure}[p]
    \centering
    \includegraphics[width=\textwidth]{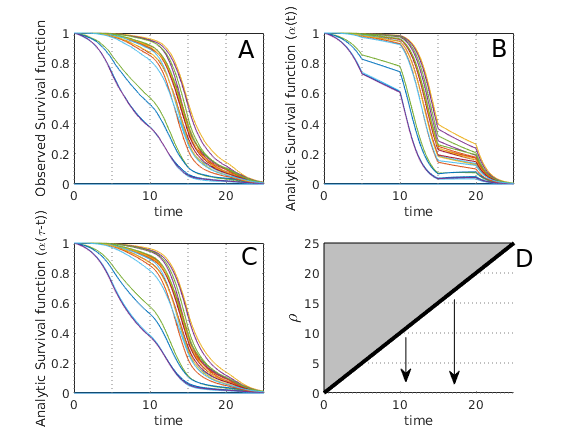}
    \caption{(A) Survival curves observed over the course of $12,000$ simulations, for piecewise constant $\alpha_k$. The values of $\alpha_k$ change when $t=5,10,15..$, indicated by the dashed grey lines. $\beta_k=0$, $\gamma=0.063$. (B) $S_a^b(t)$ as calculated using equation \ref{eq:MAIN} by simply substituting in the time varying $\alpha_k(t)$ in for $\alpha_k$. This approach violates our Markov assumption and gives non-physical results. (C) Survival curves $S_a^b(t,0)$, calculated using equation \ref{eq:MAINrho}. By correctly accounting for time variation in $\alpha_k$ we are able to match observed survival curves exactly. (D) Schematic diagram showing the flow of information in equation \ref{eq:MAINrho}; solution curves trace back from our initial conditions at $t=\rho$ to our survival curve of interest at $\rho=0$. The invalid region $t<\rho$ is greyed over.}
    \label{fig:TimeVaryParam}
\end{figure}

In order to see the importance of this Markov assumptions,
suppose we take equation \ref{eq:MAIN} and replace $\alpha$ with some piecewise constant function $\alpha(t)$; both in our simulation, and in the ODE. As can be seen in figure \ref{fig:TimeVaryParam} (top panels), simply replacing $\alpha$ with $\alpha(t)$ leads a rapid divergence of $S_a^b(t)$ and our observed survival curve. Under such naive assumptions, our calculated $S_a^b(t)$ is no longer monotone decreasing, and hence can no longer be considered a survival curve in any sense of the word. This breakdown comes about because even though $S_a^b(t)$ does not track possible states of the population directly, it must (inevitably) encode these states implicitly. When $\alpha$ changes midway through the process, this encoding is rendered invalid, and $S_a^b(t)$ no longer behaves in a sensible manner. Without the Markov assumption, $S_a^b(dt+t)$ is no longer a superposition of  $S_k^b(t)$.

Calculation of survival curves when dealing with time varying parameters is possible however. To do this, we expand the one dimensional $S_a^b(t)$ to the two dimensional $S_a^b(t,\rho)$. We define $S_a^b(t,\rho)$ to be the probability that an infected traveller, initially observed at position $a$ at time $\rho$ has no decedents arriving at $b$ before time $t$. By definition $S_a^b(t,\rho)=1$ whenever $t \le \rho$, and $S_b^b(t,t)=0$.
Using similar arguments to equation \ref{eq:preMAIN} we find that $S_a^b(t,\rho)$ is a superposition of $S_a^b(t,\rho+d\rho)$. Taking limits and converting into a differential equation we find:
\begin{equation}
    -\frac{\partial S_a^b(t,
    \rho)}{\partial \rho}= T(\rho) S_a^b(t,
    \rho) + [\beta(\rho) - \alpha(\rho) S_a^b(t,
    \rho)]\times[1-S_a^b(t,
    \rho)].
    \label{eq:MAINrho}
\end{equation}
Combining equation \ref{eq:MAINrho} with the initial conditions given at $t=\rho$, it is possible to determine $S_a^b(t,0)$ (the survival curve of interest) for any given  time $t$ by solving a simple ODE in the $\rho$ direction (see figure \ref{fig:TimeVaryParam}, left panels).
This approach accounts for time varying parameters whenever $\alpha$,$\beta$,$\gamma$ and $T$ vary independently from the epidemic course. In the case where parameters are dependent on epidemic spread (for example, border closures in response to observed spread of disease), more complicated methods are needed.
Because eq. (\ref{eq:MAINrho}) must be solved independently for each $t$ this approach is more computationally costly than the approach taken for Markov systems.

\section{Gallery}
\label{app:gallery}
\begin{figure}[p]
    \centering
    \includegraphics[width=\textwidth]{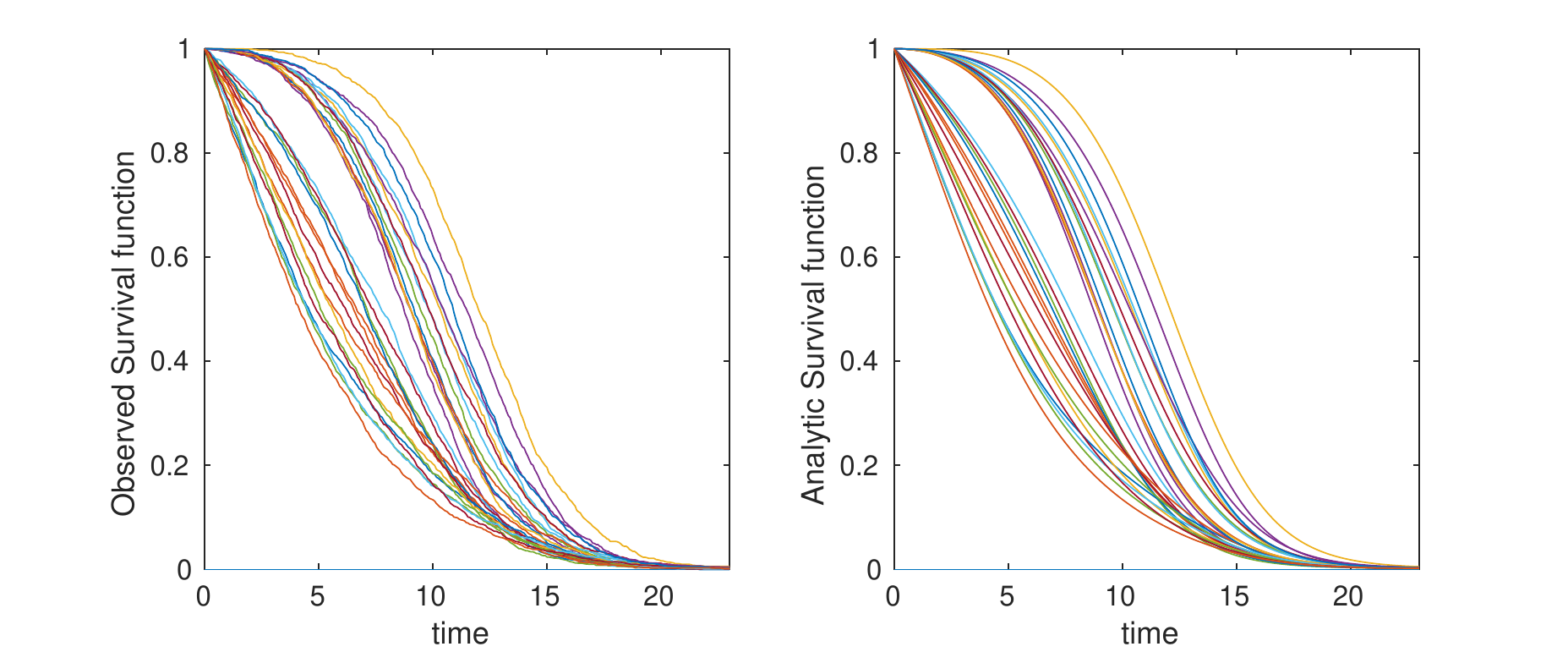}
    \caption{Here we simulate a scale free network as produced by Mathew George's SFNG function \cite{mathew_george_b-scale-free_2020}. Here we set $N=30$ nodes, infection rate $\alpha=0.5$ and recovery rate $\beta=0.1$. (Left) Observed survival curves given $1500$ simulations. Simulation done using the exact Gillespie algorithm \cite{gillespie_exact_1977}. (Right) Survival curves as calculated using eq. (\ref{eq:MAIN}). }
    \label{fig:ScaleFreeVsExperiment}
    \includegraphics[width=\textwidth*45/100]{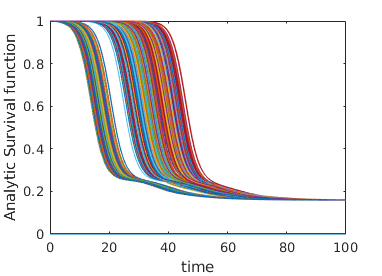}
        \includegraphics[width=\textwidth*45/100]{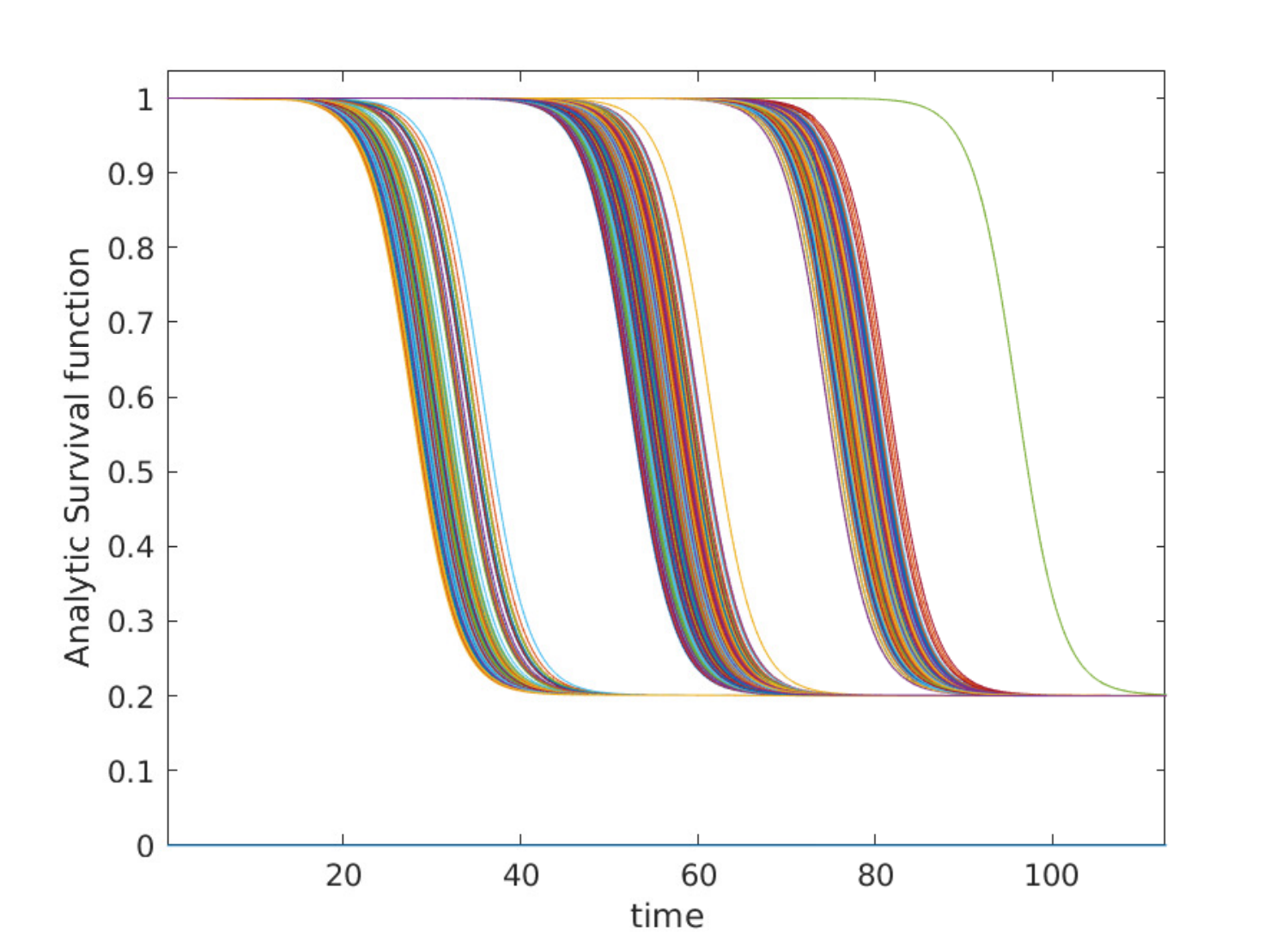}
    \caption{(Left) Survival curves as calculated using equation (\ref{eq:MAIN}), for a larger scale-free network. $N=3000,  \gamma=0.001, \alpha=0.5\pm0.11, \beta=0.1\pm0.03$. 
    (Right) Analytic Survival curves for $N=3000,  \gamma=10^{-5}, \alpha=0.5, \beta=0.1$. In both cases we observe `logistic' decay from $1$, along with infinite survival time in cases where the epidemic goes extinct, which occurs with probability $\beta_1/(\beta_1+\alpha_1)$. For small $\gamma$, time of arrival is determined predominantly by the number of steps. Here we present only analytic results, as simulations are prohibitively expensive to run. }
    \label{fig:ScaleFreeBigNetwork}
\end{figure}

\begin{figure}[p]
    \centering
    \includegraphics[width=\textwidth]{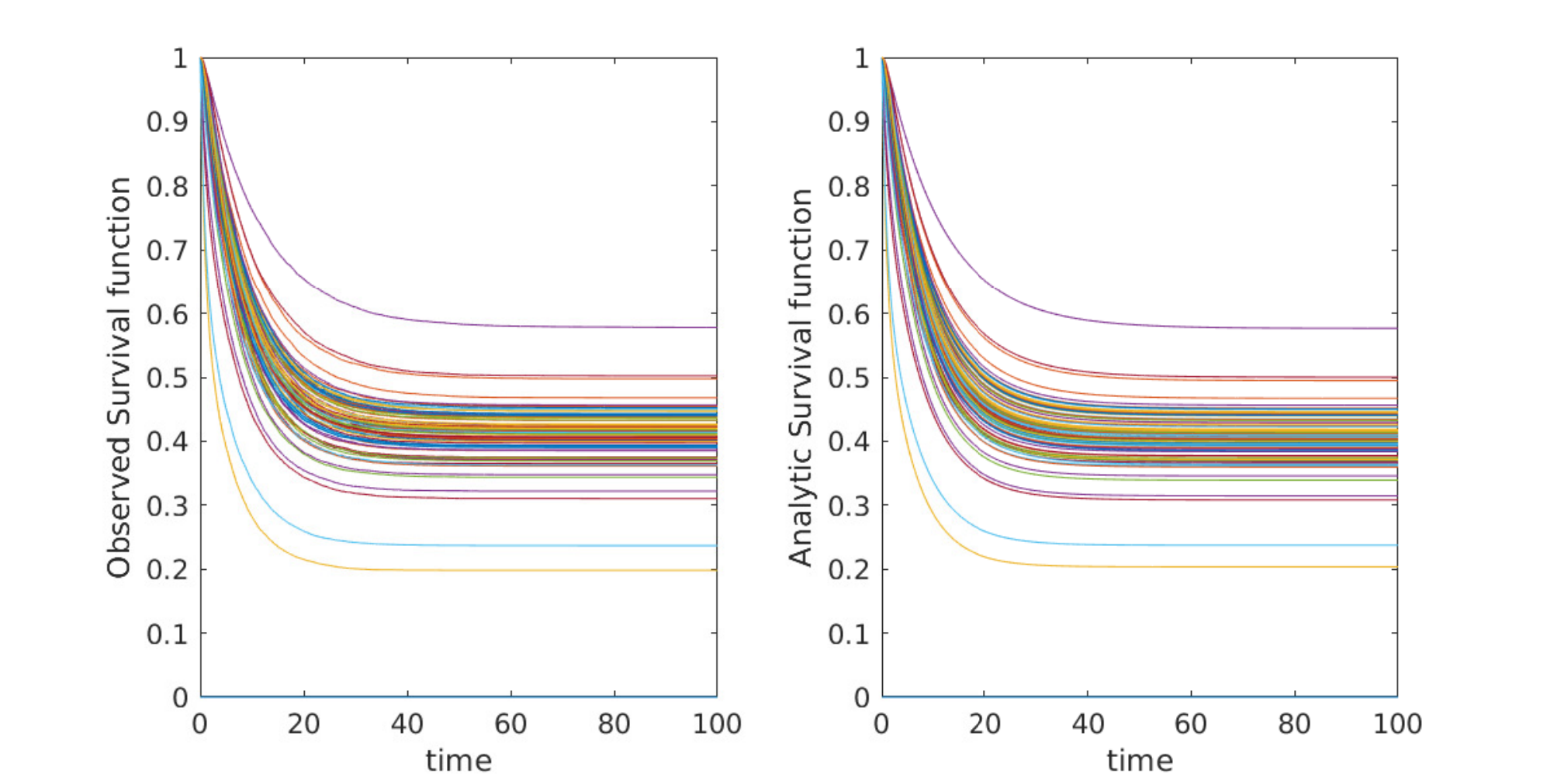}
    \caption{The analytic survival curve gives valid predictions even far from the usual parameter regime; here we consider a traveller on a random network with, $\gamma=10, \beta=0.05,\alpha=0$, that is to say, a fast moving traveller that eventually decays, but does not duplicate. Arrival time distributions as observed in $15000$ simulations perfectly match survival probabilities as calculated numerically using eq. (\ref{eq:MAIN}). Unlike the $\gamma\ll 1$ domain, where survival probabilities decay approximately logistically, here we observe roughly exponential decay. Also in contrast to the $\gamma \ll 1$ case, different locations plateau at different levels; representing different probabilities that the traveller will \emph{ever} arrive.}
    \label{fig:FastDecayingTravl}
    \includegraphics[width=\textwidth*45/100]{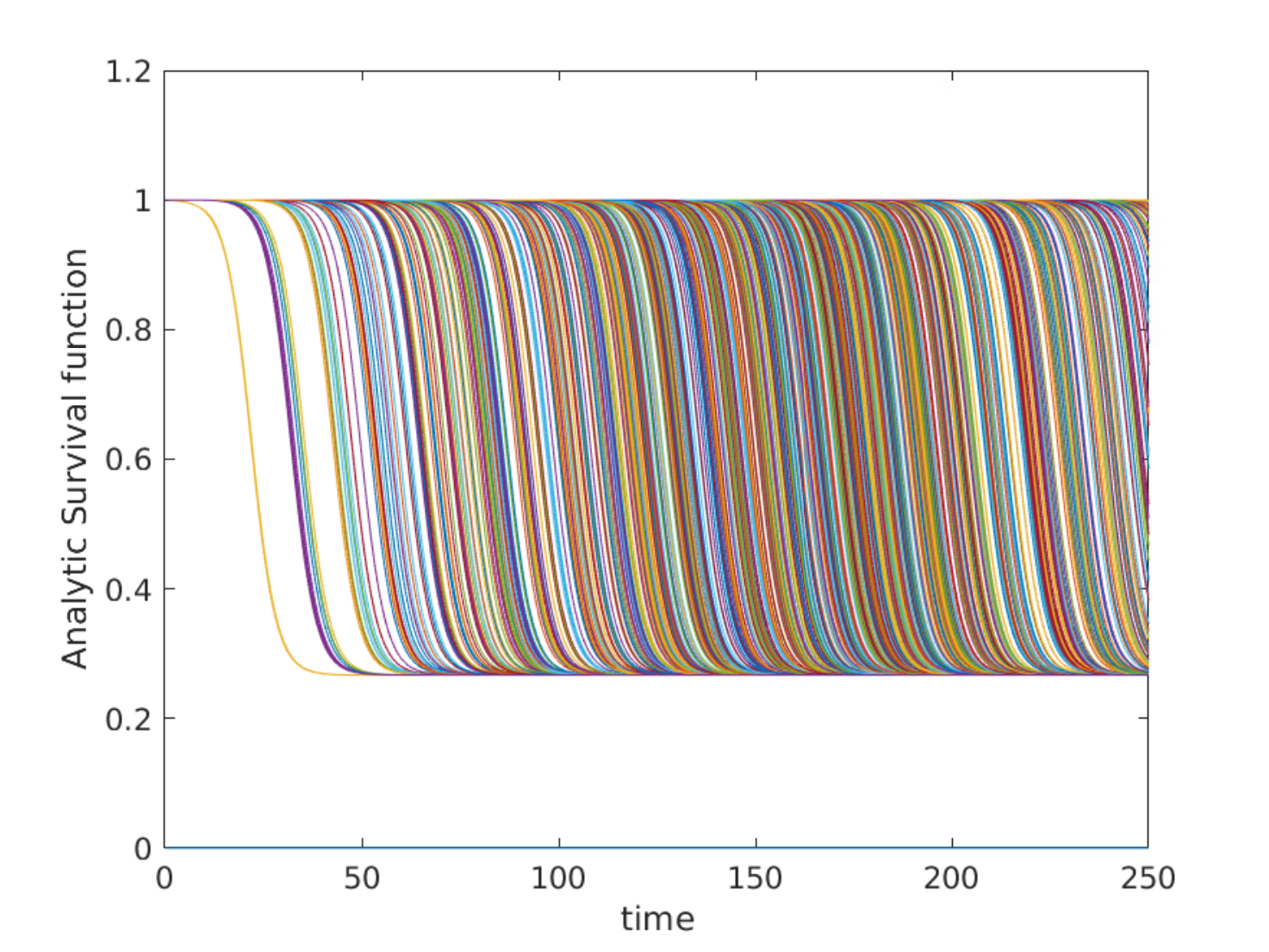}
        \includegraphics[width=\textwidth*45/100]{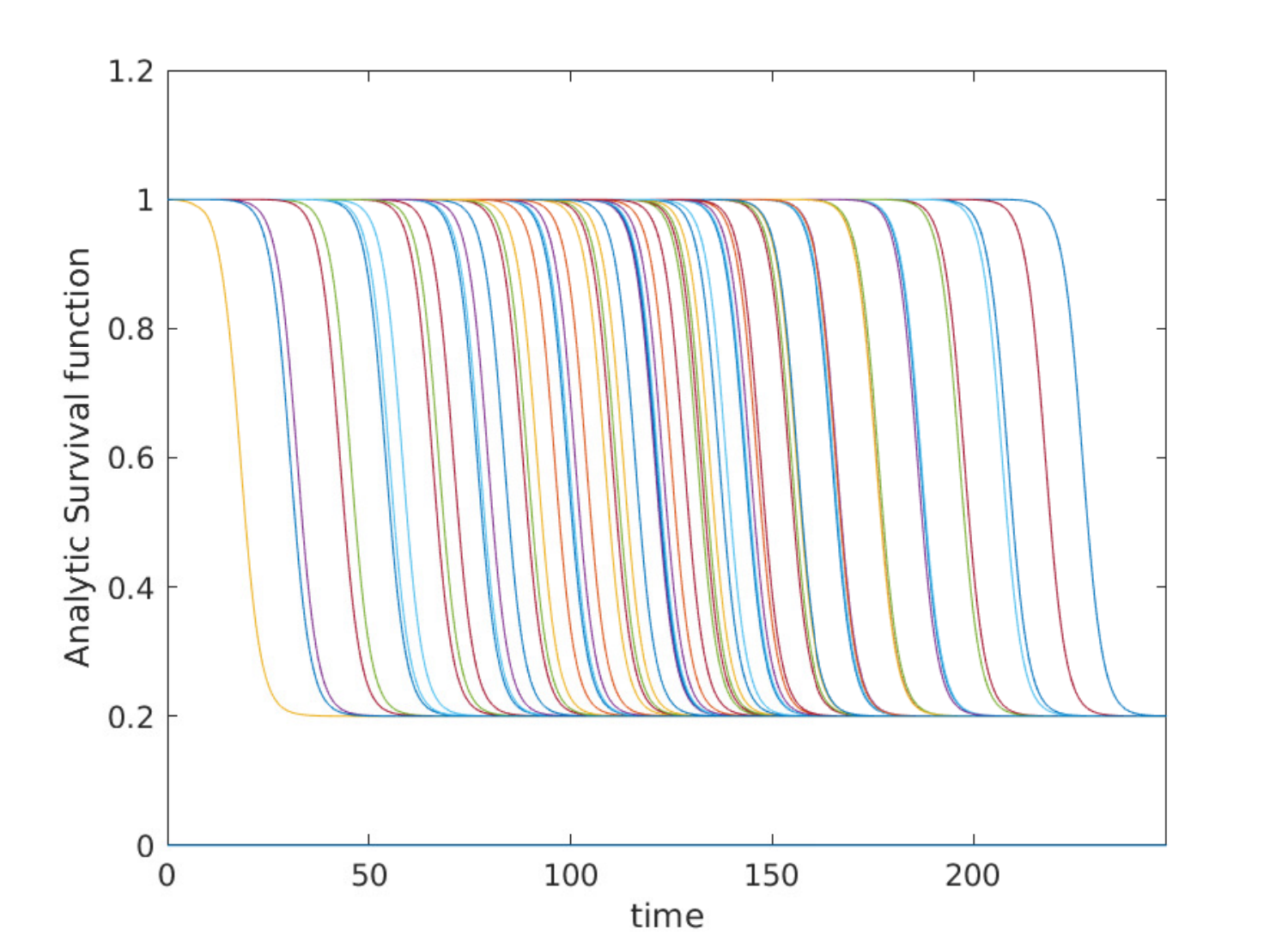}
    \caption{(Left) Analytic survival time curves, for a $30\times 30$ lattice, with periodic boundary conditions. $N=900,  \gamma=0.001$, and epidemic parameters vary by node: $\alpha=0.5\pm0.11, \beta=0.1\pm0.03$. 
    (Right) Analytic Survival curves for $20\times 20$ lattice, $N=400,  \gamma=0.001$, epidemic parameters constant across all nodes, $\alpha=0.5, \beta=0.1$. In both cases we observe `logistic' decay from $1$. Mean arrival time scales with distance from the initial site of infection. When epidemic parameters are homogeneous, symmetry results in many survival curves overlapping perfectly (the survival curve for `one step north and one step east' is the same as `one step south, one step west'). For variable parameter values, this symmetry is broken and we see a smooth distribution of arrival times.}
    \label{fig:LaticeObserved}
\end{figure}

Here we give a variety of figures, exploring the possible behavior of arrival time distributions in a variety of networks and parameter regimes.
In fig. \ref{fig:ScaleFreeVsExperiment} and \ref{fig:ScaleFreeBigNetwork} we examine a variety of scale free networks, varying network size, and parameter values. In fig. \ref{fig:FastDecayingTravl} we demonstrate the robustness of eq. (\ref{eq:MAIN}) to unusual parameter values by consider the (unrealistic) case of a fast moving traveller who is non-infectious. We observing tight agreement between analytic predictions and observed survival time in simulations.
Finally, in \ref{fig:LaticeObserved} we consider arrival times for travellers on a 30 by 30 mesh.

In all cases where simulation is feasible, we observe strong agreement in arrival probabilities between analytic and simulation based results, for all $t$ values sampled.

\section{Construction of Arbitrary Survival Curves}
\label{app:ArbitraryCurves}
One of the more concerning results of eq. (\ref{eq:MAIN}) is the implication that it is possible to construct networks with arbitrary survival curves. In this section we sketch a method for constructing such a network. The purpose of this exercise is to demonstrate that for arbitrary networks, the probability distribution of arrival times may have arbitrary complexity, and hence that any analytically tractable results require further assumptions to be made on $\alpha, \beta$ and $T$. More detail would be needed in order to make the argument rigorous, the work here is only intended as a proof of concept.

\begin{figure}[p]
    \centering
    \includegraphics[width=\textwidth]{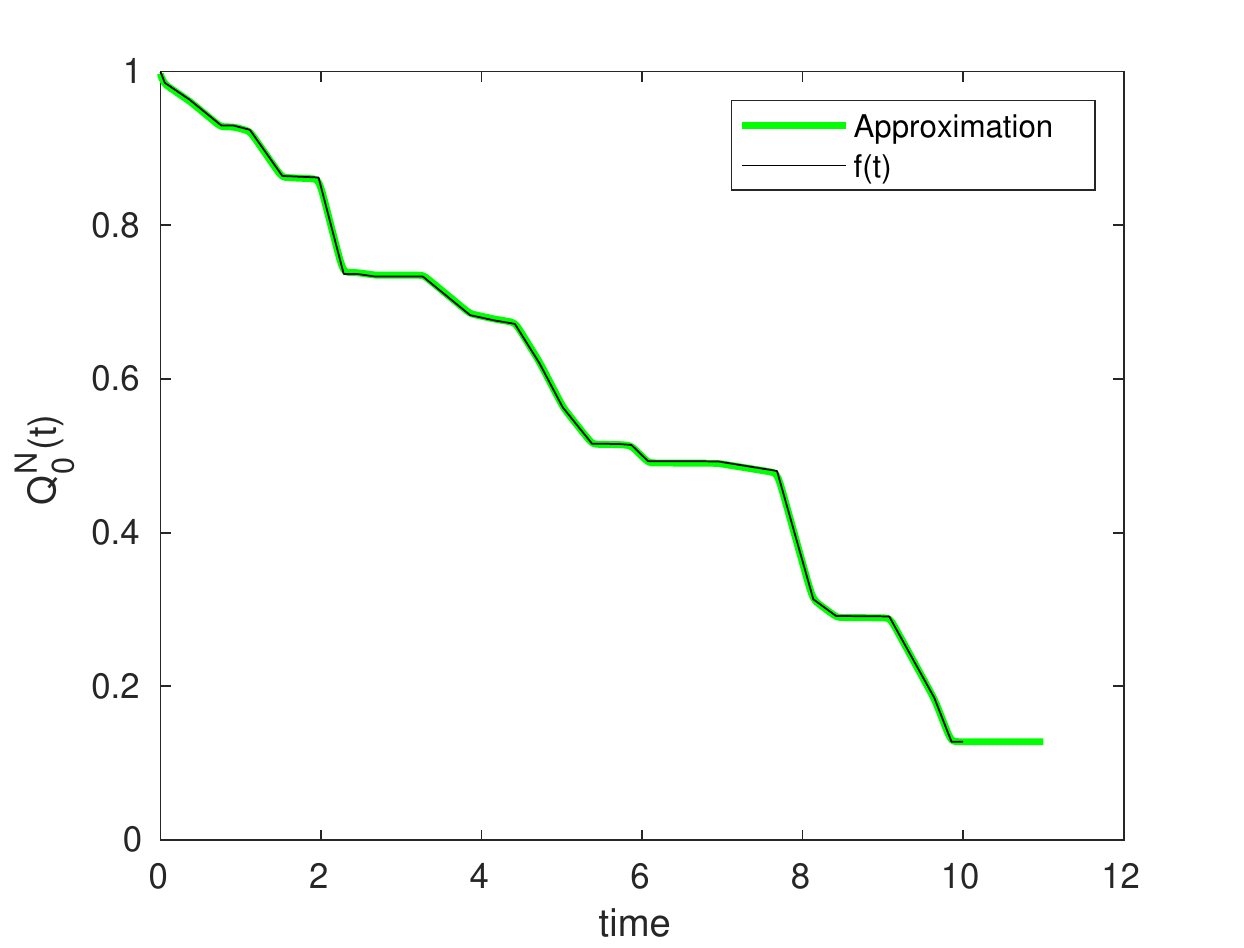}
    \centering
    \includegraphics[width=\textwidth]{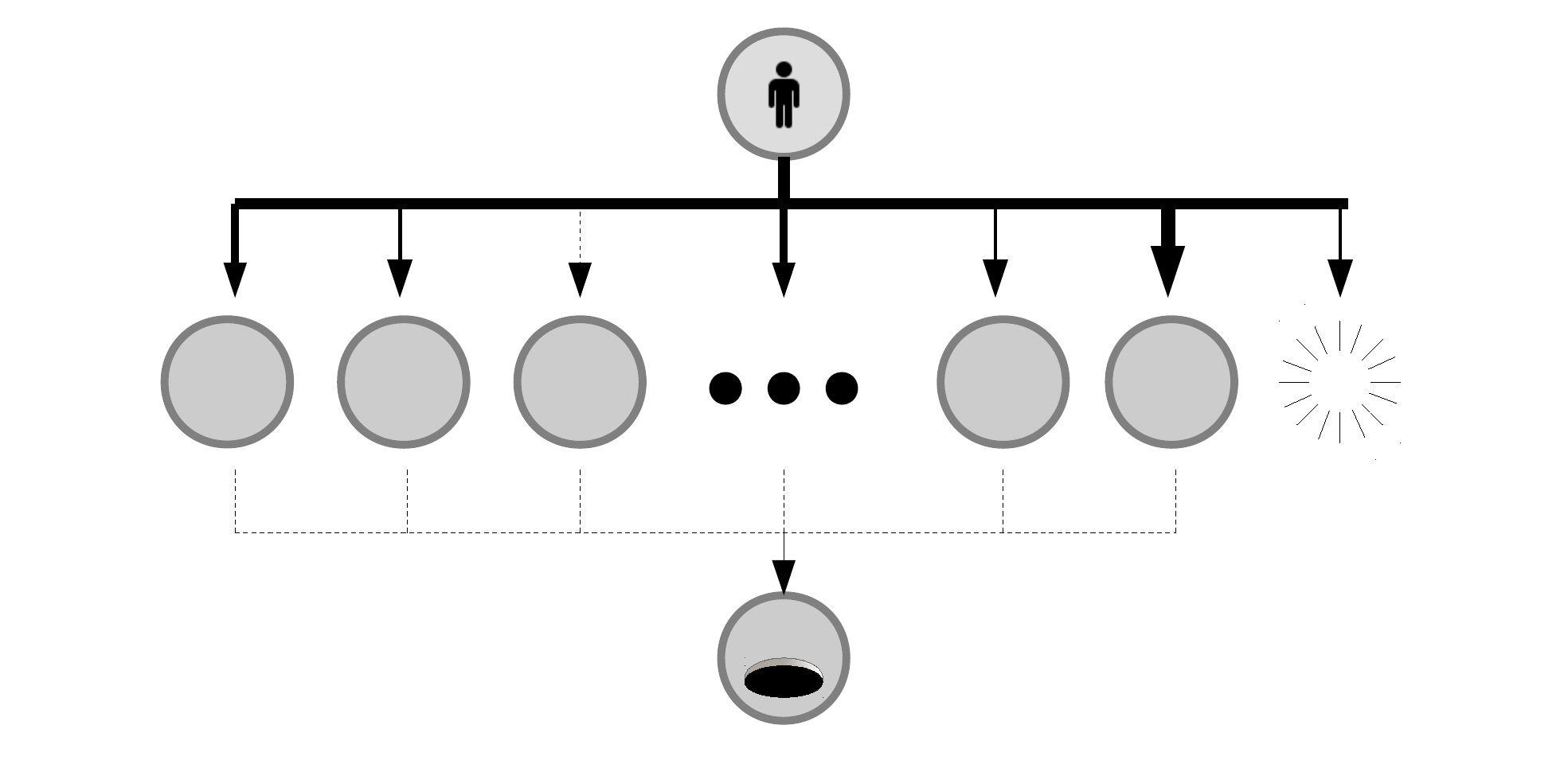}
    \caption{(Top) We generate a random piecewise linear decreasing function, and then approximate it using the survival curve of a network. Here we use \emph{analytic} expressions for our stepwise function as given in eq. (\ref{eq:logisticStep}). Direct integration of eq. (\ref{eq:MAIN}) runs into numerical difficulties due to the exceptionally small $T_{k,N}$ involved.
    (Bottom) A schematic diagram of the diamond network used to match our randomly generated survival curve. Infection rate in our central layer, along with transport rate to the `target' node is selected such that $Q_k^N(t)$ for each of our intermediary nodes approximates a step function, with a wide array of `step times'. Transport rates from our starting node to intermediate nodes are selected such that $Q_0^N(t)$ is (approximately) a superposition of these many step functions, with appropriate weight given to each such that $|Q_0^N(t)-f(t)|$ is small. }
    \label{fig:ArbitraryApprox}
\end{figure}

Suppose we are given some function $f(t)$ such that $f(0)=1, f'(t)\le 0, f(t)\ge0$. We assume that $f'(t)$ is well defined for all time. Our goal is to create a network such that $S_0^N(t)$ approximates $f(t)$ as closely as possible. Consider a simple ``diamond'' network as depicted in figure \ref{fig:ArbitraryApprox}. In this network, node zero has a directed edge connecting it to nodes $1$ to $N-1$, and nodes $1$ to $N-1$ have a single outgoing edge, directed towards the `final' node $N$. We assume $\beta_k=0$ for all intermediary nodes. We take a travel rate $\gamma=1$. 

For these ``intermediate'' nodes, the survival time is governed by the equation:
\begin{equation}
\dot S_k^N(t)=  -T_{k,N} S_k^N(t) - \alpha_k S_k^N(t)+\alpha_k [S_k^N(t)]^2,
\end{equation}
which permits solutions of the form:
\begin{equation}
S_k^N(t)= \frac{\alpha_k + T_{k,N}}{\alpha_k + (\alpha_k+2 T_{k,N})e^{(\alpha_k + T_{k,N})(t-\mu_k)} }.
\label{eq:logisticStep}
\end{equation}
Here $\mu_k$ is selected such that $(\alpha_k+2 T_{k,N})e^{-\mu_k(\alpha_k + T_{k,N})}= T_{k,N}$.
Because both $\alpha_k$ and $T_{k,N}$ are free parameters, it is possible to select them so as to construct a logistic function with arbitrary mean and scale parameters. This allows us to approximate arbitrary step functions.

We now wish to select the transition rates $T_{0,k}$ such that $S_0^N(t) \approx f(t)$. We select $\alpha_0=0$ such that the initial infection at node zero does not replicate, and instead simply `jumps' to one of our $N-1$ intermediary nodes, or recovers. 
$S_0^N(t)$ is governed by the equation:
\begin{equation}
\dot S_0^N(t)=  \sum T_{0,k} \left[S_k^N(t)-S_0^N(t)\right]  \beta_k \left[1-S_0^N(t)\right]
\end{equation}
If we select $T_{0,k}$ and $\beta_k$ very large, then our initial infection will linger in the starting node for only a short time, and 
\begin{equation}
S_0^N(t)\approx  \frac{\sum T_{0,k} S_k^N(t) + \beta_k}{\sum T_{0,k} + \beta_k},
\end{equation}

That is to say, $S_0^N(t)$ is a sum of step functions, and a constant.
Transition parameters are selected such that $\beta_k/(\sum T_{0,k} + \beta_k)$ is equal to $f(\infty)$ and  $T_{0,k}/(\sum T_{0,k} + \beta_k)= f( (\mu_k+\mu_{k+1})/2)-f( (\mu_k+\mu_{k-1})/2)$. In this way $f(t)$ can be approximated using a series of (approximate) step functions.
A demonstration of this is given in figure \ref{fig:ArbitraryApprox}. More sophisticated algorithms would presumably be able to approximate $f$ using fewer nodes, but for our purposes this simple approach will suffice.

\section{Survival curves for non-zero $\beta$}
\label{app:BetaNonZero}
Throughout the main text, we frequently made use of the simplifying assumption $\beta=0$; an unrealistic assumption in real world context.
Suppose we wish to consider non-zero $\beta$, in the context of global epidemic spread. We assume, as previously, that travel is rare, and also that we interest ourselves only in diseases such that $R_0>1$, hence $\gamma \ll \beta < \alpha$.
This leads to the equation
\begin{align}
    \dot {S}^{b}_a(t) \approx& \left(\beta_a-\alpha_a {S}^{b}_a(t) \right)\left(1- {S}^{b}_a(t) \right),
    \intertext{and hence,}
    {S}^{b}_a(t) \approx & \frac{1 + \beta_a e^{(\alpha_a -\beta_a)(t-\mu_a^b)}}{1 + \alpha_a e^{(\alpha_a -\beta_a)(t-\mu_a^b)}},
    \label{eq:logistic}
\end{align}
that is to say, in the limit of small $\gamma$, the arrival time distribution can be approximated using a logistic function.
As in the main text $\mu_a^b$ is some mean arrival time constant determined by $a$,$b$, and $\gamma T$. In the limit $t\rightarrow \infty$ ${S}^{b}_a(t)\rightarrow \beta_a/\alpha_a$, that is to say, the probability that $\tau_a^b=\infty$ is precisely the probability of early epidemic extinction in our branching process. Unfortunately, this non-zero probability of `infinite' arrival time leads to $E(\tau_a^b)=\infty$; not an especially informative result. It is thus useful to instead consider the arrival time \emph{conditional} on $\tau_a^b<\infty$:
\begin{align}
    {S}^{b}_a(t) =& \frac{{S}^{b}_a(t)-{S}^{b}_a(\infty)}{{S}^{b}_a(0)-{S}^{b}_a(\infty)}, \\
    \approx&
    \frac{1}{1 + e^{(\alpha_a -\beta_a)(t-\mu_b)}}.
    \label{eq:ConditionalSurvCurv}
\end{align}
At this stage we have an `ideal' logistic distribution \cite{balakrishnan_handbook_2013}, with mean value $\mu_a^b$ (currently unknown), and scale parameter $1/(\alpha_a -\beta_a)$. Hence, when making predictions concerning real world epidemics that have not gone extinct in their early stages, it is useful to remap our variables such that we imagine $\beta_k=0$ and $\alpha_k$ is the \emph{net} grow rate of our disease (formerly labeled $\alpha_k-\beta_k$).

\section{Comparison to, and improvement upon past exponential growth methods.}
\label{app:chen}
In this appendix we will discuss the Matrix Exponential method, as used by Chen et al. \cite{chen_estimating_2018} (referred to by them as `linear spreading theory'). We begin by introducing the method itself, and its connection to the Branching Process approach studied in the main text. We then explain some of the computational details necessary for fast calculation, and the circumstances under which the approximation is and is not accurate.

Let us begin by describing the method itself. Suppose, at time $t=0$, we have a population of infected individuals ${\bf P}(0)$ -- that is to say, there are $P_k(0)$ infected individuals located at node $k$. These individuals travel between nodes at transport rate $\gamma T$, and infect others at some rate $\alpha$. This rate is assumed to be constant; we assume that the total population is large enough such that population constraints are not relevant on the timescale we are interested in.
Under this unbound growth assumption, the {\it expected} number of infectious individuals if governed by the linear equation:
\begin{align}
    \dot {\bf P}(t) =& \gamma T^\top {\bf P}(t) + \alpha {\bf P}(t). \\ 
    \intertext{This equation permits solutions of the form}
    {\bf P}(t) =& \exp[ (\gamma T^\top + \alpha I)t] {\bf P}(0).
\end{align}
Here, we make use of the matrix exponential; that is to say $\exp M = I +M + M^2/2! + M^3/3!+...$ .
Here we use the transposed transition matrix, $T^\top$, because here we consider the flow of infected individuals {\it forward} through the network, as opposed to the flow of arrival probability backwards through the network (as we do in the main text).

Suppose we start with an initial population of $1$ in node $a$. If we wish calculate the deterministic time when the \emph{expected} population in node $b$ reaches $1$, we must solve ${\bf P}_b(t)=1$. Written slightly differently,
\begin{align}
    1 =&  \delta_b^\top \exp[ (\gamma T^\top + \alpha I)t] \delta_a, \label{eq:ChenTranscendent}
\end{align}
where $\delta_a,\delta_b$ are indicator vectors, equal to $1$ at index $a$ and $b$ respectively, and equal to zero elsewhere. 
This equation is equivalent to equation 7 of Chen et al \cite{chen_estimating_2018}. Here we have reached the equation via a direct appeal to unbound population growth, while Chen et al instead describe their formalism as a linearisation of a full SIR model.\\
Eq. (\ref{eq:ChenTranscendent}) can also be reached via an appeal to branching processes. By making the change of variables ${S} \rightarrow {Q}$ as we do in the main text, suitable approximations lead to:
\begin{align}
  {Q}^{b}_a(t) \approx&  e^{(\gamma T + \alpha I)t} {Q}_a^b(0),
\end{align}
eq. (\ref{eq:MatrixExpQ}) of the main text. The expected arrival time at node $a$ is precisely the time when ${Q}^{b}_a(t)=1$, that is to say 
\begin{align}
  1&= \delta_a^\top  e^{(\gamma T + \alpha I)t} \delta_b.
  \label{eq:IterGoal}
\end{align}

The two methods give the same results (under the influence of suitably powerful approximations).

Before moving on to discuss the accuracy of these results, let us first discuss some computational difficulties. Eq. (\ref{eq:ChenTranscendent}) is transcendental and can thus only be solved approximately, using some suitable numerical scheme. In their paper, Chen et al. approach this difficulty by identifying the minimal number of steps $d$ required to get from $a$ to $b$, and truncating the polynomial expansion of their matrix exponential after this many steps:

\begin{align}
  \delta_b^\top e^{(\gamma T^\top + \alpha I)t} \delta_a =& e^{\alpha t} [ \delta_b^\top \delta_a  + \delta_b^\top \gamma T^\top t \delta_a + ...+\delta_b^\top \frac{\gamma^d T^{\top d} t^d}{d!} \delta_a+...  ] \\
  1 \approx&  e^{\alpha t} \delta_b^\top \frac{\gamma^d T^{\top d} t^d}{d!} \delta_a.
  \label{eq:invLambertW}
\end{align}

All terms before the $d^{th}$ term are zero, all terms after are higher powers of $\gamma$ and thus assumed to be small. Chen et al solve eq. (\ref{eq:invLambertW}) using the Lambert-W function \cite{leonhard_euler_serie_1783,weisstein_lambert_2020}, a nonlinear function originally constructed to solve equations of the form $y e^y =x$. In cases where we have a single ``most probable'' path from $a$ to $b$, we can make the approximation:
\begin{align}
  \alpha t \approx&  -\log \left( \delta_b^\top \frac{\gamma^d T^{\top d} t^d}{d!} \delta_a \right),\\
  \approx& -d \log{\gamma} - \log \left(\prod T_{i,j} \right),\\
  \approx& - \sum \log{\gamma} + \log  T_{i,j}.
  \end{align}
Here $T_{i,j}$ are the transition rates along the steps in our shortest path. Such an estimate neglects terms of magnitude $d \log(t/d)$, along with contributions from all paths except the shortest. This result mirrors those of both Gautreau et al. \cite{gautreau_arrival_2007}, as well as Brockmann et al. \cite{brockmann_hidden_2013} (with the notable difference being that $-\log \gamma$ is replaced by $1$ when using the effective distance metric).

An alternative method to solve Eq (\ref{eq:ChenTranscendent}) (not used by Chen et al) is the iteration scheme:
\begin{align}
1=& \delta_a^\top  e^{\gamma T t_n} e^{\alpha t_{n+1}} \delta_b,\\
\alpha t_{n+1}=& -\log\left(\delta_a^\top  e^{\gamma T t_n} \delta_b \right).
\label{eq:iteration}
\end{align}
Here we make use of the fact that $\alpha t_{n+1}$ varies faster than $\gamma T t_n$; hence, if we have even a moderately accurate approximation $t_n$, $e^{\gamma T t_n}$ will be close to the correct value $e^{\gamma T t}$. In contrast $e^{\alpha t_{n+1}}$ varies quickly: demanding $t_{n+1}$ to solve eq (\ref{eq:iteration}), we quickly approach true solutions of eq. (\ref{eq:IterGoal}).

This iteration scheme requires us to calculate large matrix exponentials repeatedly, an operation which is generically computationally expensive, but can be made significantly less costly by first computing the eigenvector decomposition $T = V D V^{-1}$. Here $V$ is a matrix containing the eigenvectors of $M$, while $D$ is a diagonal matrix containing the corresponding eigenvalues. This allows us to instead compute the matrix exponential via elementwise exponentials of the diagonal elements:
\begin{align}
\alpha t_{n+1}= -\log\left( (\delta_a^\top  V) e^{\gamma D t_n} (V^{-1}\delta_b) \right)
\end{align}

Matlab code to implement this iteration scheme, approximating transport times \emph{to} node $b$ from all starting points is given by:

\begin{verbatim}
%%Set up
[V,D_original]  =eig(T);
N=size(T,1);
b=7; %Example start/end points.
delta_b= zeros(N,1);
delta_b(b)=1; %%Set up our initial vectors.

D= gamma * diag(D_original); %D is a vector of length N containing eigenvalues.
RightVect= V\delta_b;

Tk=-ones(N,1);

for(aaa=1:N)
    guessT= (-log(gamma))/alpha
    deltaT=5;
    LeftVect= V(aaa,:);
    while(abs(deltaT)>10^-3)
         expGammaT= LeftVect*(exp(D*guessT).*RightVect);
         deltaT= -log(expGammaT)/alpha
    end
    Tk(aaa)=guessT;
end
\end{verbatim}

The computational cost of the above code is overwhelmingly dominated by \verb|eig(M)|, and hence the runtime scales like $O(n^3)$, similar to the Floyd–Warshall algorithm as might be used to identify shortest paths when calculating the effective distance. 
It is possible to calculate the time taken to arrive in each location \emph{from} a fixed starting location $a$ by using the transposed transition matrix, $T^\top$, rather than $T$. 

Now, in all uses of such methods, it is important to note that the deterministic time when the \emph{expected} population reaches $1$ (as calculated using Chen et al's method), and the \emph{expected} time when the stochastically varying population reaches one are \emph{not} equivalent concepts. This is best illustrated by considering the extreme case $\alpha=\beta=0$, where we observe that $\delta_a^\top  e^{\gamma T t} e^{\alpha t} \delta_b<1$ for \emph{all} $t$. Taken at face value this would erroneously imply that the arrival time has an expected value of infinity; in practice this merely illustrates the difficulties of using such an approximation for diseases with low $\alpha$ (such as HIV).
In general, eq. (\ref{eq:MatrixExpQ}) is observed to be a good approximation when $\alpha \gg \gamma T$, and performs poorly when $\alpha \ll 1$.

\section{Variance in Arrival Time}
\label{app:varianceHighPop}
Here we provide the algebraic details for calculating variances in arrival time, assuming a starting population greater than 1. 

Recall that $\tau_a^b$ denotes the arrival time at $b$, assuming an epidemic starts at $a$, and has mean $\mu_a^b$. The survival function of $S_a^b(t) = P(\tau_a^b>t)$. In what follows, we suppress subscripts and superscripts, considering some generic $\tau$.

\begin{align}
E(\tau)= \int_0^\infty S dt&= \mu \\ \text{  Assuming the mean}& \text{ value $\mu \gg 0$} \nonumber\\
\dot S&= \alpha S(1- S)\\
\int_0^\infty S^k - S^{k+1} dt &= \int_0^\infty S^{k-1} \dot S/\alpha dt \\
\int_1^0 \hat S^{k-1}/\alpha dS &= \left[\frac{S^k}{\alpha k} \right]_1^0 = -1/\alpha k\\
\implies \int_0^\infty S^{k+1} dt&=  \int_0^\infty \hat S^{k} dt - \frac{1}{\alpha k} = \mu - \sum_{k=1}^N \frac{1}{\alpha k}
\label{eq:ExpectationByPopRecursion}
\end{align}
%\todo{check for sign errors}
That is to say, if the expected arrival given an initial population of $1$ is $\mu$, then the arrival time assuming $K+1$ individuals is $\mu - \sum_1^N 1/\alpha k$. This makes sense, given that the expected time of transition form $k$ to $k+1$ individuals, for a branching rate of $\alpha$ is precisely equal to $1/\alpha k$.

Next, we need to determine the expected value of $E(\tau^2)$ for a starting population of $N$.
For a starting population of 1, we have $E(\tau^2) = \mu^2 +  \pi^2/3 \alpha^2$ \cite{balakrishnan_handbook_2013}. For larger populations:
\begin{align}
    \int_0^\infty 2t (S^k - S^{k+1}) dt &= \int_0^\infty 2t S^{k-1} \dot {S} /\alpha dt \\
    S = 1/(1+\exp[\alpha (t-\mu)]) &\implies t= \log\left(\frac{1-S}{S} \right)/\alpha +\mu \\
    \frac{2}{\alpha^2}\int_1^0 \log\left(\frac{1-S}{S} \right) S^{k-1} d S&= \frac{2}{\alpha^2}\frac{H_{k-1}}{k}\\ \nonumber \text{ where $H_k$ is the } &k^{th} \text{ harmonic number} \sum_{i=1}^k \frac{1}{i} \text{ and } H_0=0.\\
    \frac{2}{\alpha} \int_0^\infty t S^{k-1} d{S} &= \frac{2}{\alpha^2}\frac{H_{k-1}}{k} - \frac{2 \mu}{\alpha k}\\
     \int_0^\infty 2t S^{N+1} dt&= \mu^2 + \frac{\pi^2}{3 \alpha^2} + \sum_{k=1}^{N} \left[\frac{2}{\alpha^2}\frac{H_{k-1}}{k} - \frac{2 \mu}{\alpha k} \right] 
     \label{eq:varianceByPopRecursion}
\end{align}

The variance is thus given as:
\begin{align}
E(\tau^2) -E(\tau)^2&= \mu^2 + \frac{\pi^2}{3 s^2} + \sum_{k=1}^{N} \left[\frac{2}{s^2}\frac{H_{k-1}}{k} - \frac{2 \mu}{s k} \right] 
     - \left(\mu_a^b - \sum_1^N 1/sk \right)^2\\
     &= \frac{\pi^2}{3 s^2} + \sum_{k=1}^{N} \left[\frac{2}{s^2}\frac{H_{k-1}}{k}  \right] 
     - \left(\sum_1^N 1/sk \right)^2\\
     &= \frac{\pi^2}{3 s^2} + \frac{1}{s^2} \sum_{k=1}^{N} \left[\frac{2 H_{k-1}- H_N}{k}  \right] = \frac{\pi^2}{3 s^2} - \sum \frac{1}{s^2 k^2}
     \label{eq:varianceVsPopRepeat}
\end{align}

\bibliographystyle{apalike}
\bibliography{references}
\end{document}